# CHAPTER

# III-V Solar Cells

*James P. Connolly[1*], Denis Mencaraglia[2]*

[1] Nanophotonic Technology Centre, Universidad Politécnica de Valencia, Camino de Vera s/n, 46022 Valencia , Spain

[2] Laboratoire de Génie Électrique de Paris, LGEP, UMR 8507 CNRS-Supélec, Université Pierre et Marie Curie, Université Paris-Sud, 11 rue Joliot-Curie, Plateau de Moulon, 91192 Gif-sur-Yvette Cedex, France

[*] Corresponding author . Email : connolly@ntc.upv.es, Tel. : +34  96 387 7000 extn. 88101

**Table of Contents**

X.1 Introduction

Upward trends in energy costs are a powerful motor in the development of new energy sources, and reduce relative costs of a range of technologies. This is the case for the III-V semiconductor compounds which are traditionally[1] an expensive photovoltaic technology whilst also being the most efficient, with corresponding advantages and disadvantages.

Principal among the disadvantages are relatively complex synthesis and device fabrication, and corollary issues such as availability of relatively rare elements (In, Ga).[2] These two points are largely responsible for the higher cost.

Among the advantages, on the other hand, are a number of materials characteristics which help make III-V solar cells the most efficient photovoltaic materials available at present. The principal reason for this is the flexible combination of a range of materials from binary to quaternary compounds with a corresponding flexibility of bandgap engineering. More significantly, a number of these compounds interact strongly with light, since they largely retain direct bandgaps and correspondingly high absorption coefficients, and therefore also tend to radiate light efficiently. This is a class of materials that therefore features most of the opto-electronically efficient semiconductors.

With these advantages, the III-V semiconductors are a flexible group of materials well suited for opto-electronic applications. They are therefore good materials for high efficiency solar cells using the basic single junction concepts developed since the early days of photovoltaics. The bandgap engineering aspect allows this class of cells to be tailored to different spectra, for example global, direct concentrated or space spectra and their corresponding applications. Moreover, they are ideally suited for the fundamental development of new concepts because the flexible bandgap engineering properties allow new designs to be investigated.

The overall result of these considerations is that niche applications requiring high efficiency or fundamental research have largely driven III-V photovoltaic development to date. Historically, the first and most important niche application has involved space applications where low weight and hence high efficiency combined with reliability is the prime concern. This is currently in the process of being supplemented by terrestrial applications using the cost reducing solar concentrator technologies.

The following sections address some materials aspects of III-V solar cell development with a focus on design for maximum efficiency resting on the flexibility afforded by this family of materials.

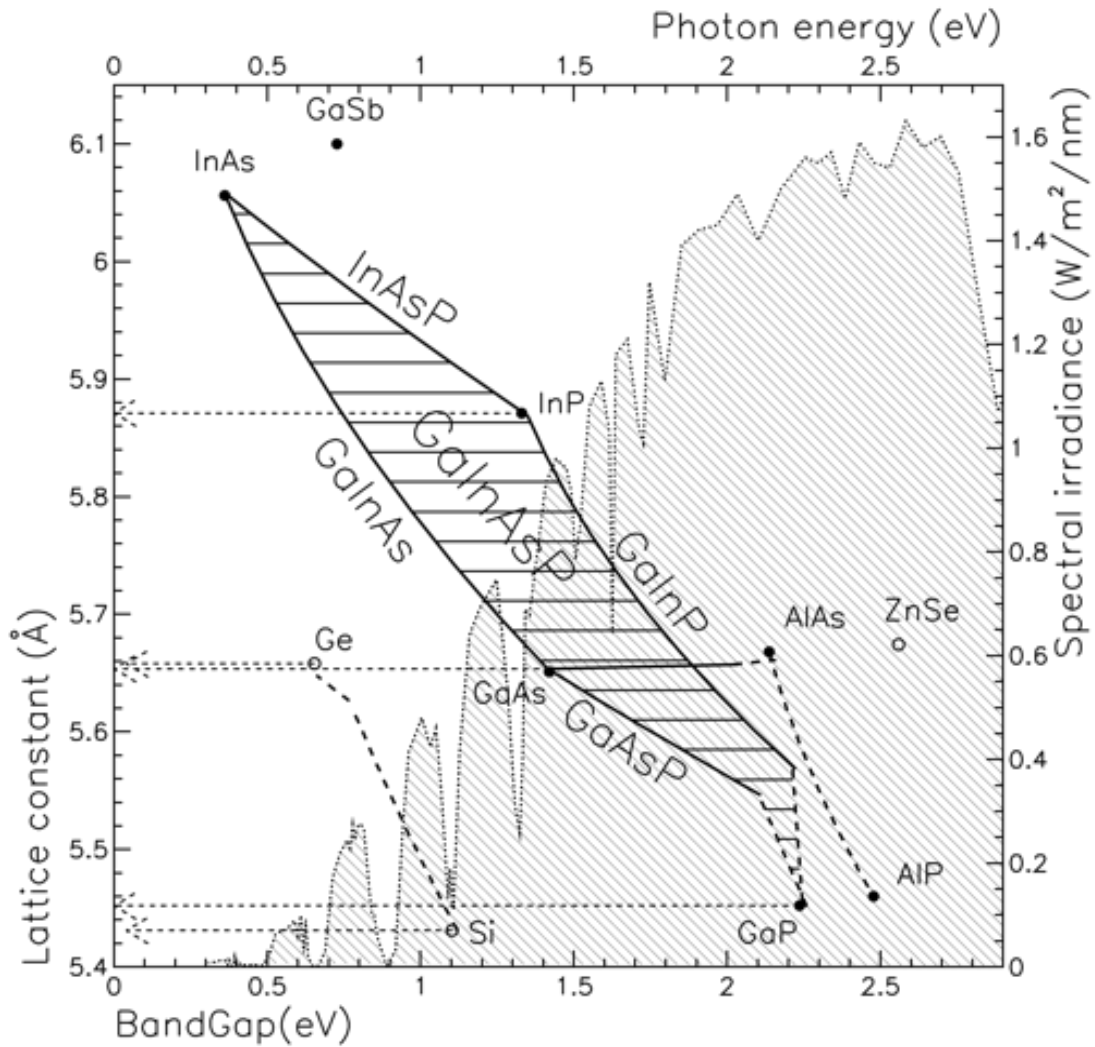

Figure 1 Significant III-V semiconductors in terms of 300K lattice parameters and bandgaps, with the horizontal dashed area indicating the range of GaInAsP compositions, and situating common III-V and group IV substrates in the context of the solar spectrum (shaded area upper, right axes).

X.2 Materials and growth

X.2.1 The III-V semiconductors

The III-V semiconductors are based on group III or Boron group and group V or Nitrogen group elements as illustrated in figure 1. This common diagramme,[3] situates the most interesting compounds for photovoltaic applications in terms of their bandgaps and lattice constants as reported in a range of sources.[4-7] Also shown is the terrestrial solar spectrum AM1.5 allowing us to see the III-V materials in the context of available power, and demonstrating good coverage of nearly the entire terrestrial spectrum.

The most commonly used substrates are first GaAs and then InP, which incidentally possess bandgaps near the ideal for solar conversion as we will see. Furthermore, materials compatible with these substrates are the most technologically important. These are first of all lattice matched compounds, which can be grown without strain relaxation and associated defects reducing device performance. The dashed arrows on figure 1 indicate substrate lattice parameters and we include here the two main group IV semiconductors Ge and Si, which as we will see provide possible routes to lower cost fabrication via heterogeneous growth. We note in passing that Ge is close to lattice match with GaAs, and that a small amount of In to GaAs can allow exact lattice matching to low cost Ge substrates. The substrate is therefore frequently used in triple designs as the lowest bandgap component as we will see subsequently.

In addition to the binary compounds, a rich family of ternary and quaternary compounds are available by substitution of a range of fractions of different group III and group V atoms while keeping a stochiometric III-V ratio.

Considering first the GaAs materials family, the ternary compound, $Al_xGa_{1-x}As$ is historically the first extensively studied material.[8-10] It retains a direct bandgap greater than GaAs over the greater part of the compositional range, and remains nearly lattice matched to GaAs across the entire range of Al fractions (fig. 1). Residual strain can be essentially eliminated with the addition of a small amount of phosphorus allowing lattice matched $AlAs_{0.96}P_{0.04}$ growth on GaAs. It suffers however from materials issues related primarily to Al and associated recombination DX centres[9-11] which limit device efficiencies unacceptably. Furthermore, these compounds are increasingly unstable and highly reactive with increasing Al composition. As a result, despite the seemingly ideal band-gap engineering potential of these compounds, their use in photovoltaic cells is largely limited to window layers, tunnel junctions, and some work on heterojunction AlGaAs/GaAs concepts.[12]

Nevertheless, there exist other less reactive Al ternary and quaternary compounds with gaps greater than GaAs. The chief amongst these for photovoltaic applications is AlInP lattice matched to GaAs. It is an important window layer and usually preferably to AlGaAs due to its lower reactivity.

Considering phosphides, the ternary compound $Ga_{0.515}In_{0.485}P$ is also lattice matched to GaAs. The gap of this material varies as a function of sublattice ordering: an ordered group III lattice yields a direct gap of 1.96eV, which is reduced, depending on the degree of disorder, by up to 0.5eV.[13]

For bandgaps lower than GaAs, there is a shortage of attractive III-V compounds compatible with GaAs. The quaternary solution $In_{1-x}Ga_xN_yAs_{1-y}$ nitrides has been proposed[14], as the addition of just a few percent of nitrogen allows lattice matching and an ideal third junction bandgap. It also, however, introduces crippling materials defects that lead to unacceptably short minority carrier lifetimes for reasons that are fully understood although insterstitial nitrogen has been shown to play a role[15]. But despite slow progress for some time, a breakthrough has recently been achieved by Sabnis *et al.* of Solar Junction, with the pentenary GaInNAsSb. They have reported[16] an independantly verified world record efficiency of 43.5% at 400 suns. Despite this impressive result, details are unavailable and the performance of this novel pentenary dilute nitride is ill defined.

Concerning substrates of InP, only the ternary $In_{0.53}Ga_{0.47}As$ of bandgap approximately 0.72eV is lattice matched to it. As we will see this is a non-ideal bandgap combination for multijunction designs. More fundamentally, InP, despite its near ideal bandstructure and corresponding limiting efficiency of 31% has achieved just 22% two decades ago with no certified progress since,[17] although related work[18] continues on cells lattice matched to InP substrates. This is due partly to inherent performance issues, and to the fact that InP is a relatively dense and rather brittle material[19,20] and therefore poses handling difficulties making industrial low cost development challenging.

The overall conclusion is that the quaternary compound GaInAsP is currently the most important materials family, including as it does compounds lattice matched to all the major substrates in use. It comprises as subsets the three important ternary phosphides which are GaAsP, GaInP, and InAsP, as well as the all important GaInAs materials family, essential in a wide range of applications.

This over-view of materials leads us to the conclusion that compounds of the GaInAsP family on GaAs and Ge substrates are the most promising. The following sections give an overview of some progress in the development of designs based on these materials.

X.2.2 Growth methods

A brief mention of III-V growth[21] is key to understanding the cost of these materials. To start with, wafer growth is by standard single crystal boule fabrication usually by one of two methods.

The first is the Czochralski method, where a single crystal seed of the material in a known crystal orientation is placed in contact with the melt comprising of a liquid solution of the same material, which may be encapsulated to prevent the evaporation of some species, and in particular As. The crystal is pulled slowly from the melt producing a single crystal ingot or boule.

The similar float zone and Bridgman alternatives consist of moving the melt away from the seed, rather than pulling the seed and crystal away from the melt. The recrystallisation occurs behind the moving heater and associated melt zone, resulting again in a single crystal boule. This method has one advantage in that impurities are expelled from the melt at the interface with the crystal. By this means, very high purity crystals can be achieved by repeated passes with the impurities segregated in the section of the boule furthest from the seed.

The boule is subsequently mechanically cut into wafers. Following this, surface treatment such as polishing produces the final single crystal wafers ready for further processing. The processing may consist of direct conversion into devices. For solar cells, the process consists primarily of diffusion or ion implantation of doping profiles defining the junction and enabling photovoltaic action, followed by additional essential features such as metal contacting which is an art we won't explore further here.

The steps described so far allow fabrication of single junction solar cells. More sophisticated structures are made using such wafers as growth substrates by further epitaxial growth techniques.

The first class is the relatively low cost chemical vapour deposition methods. The dominant variant is atmospheric pressure metal organic vapour phase epitaxy (MOVPE or MOCVD). This is a technique whereby metalorganic precursor gases, optionally including dopant species, are flowed through a growth chamber. The precursors impinge on the wafer, placed on a temperature controlled stage, leading to epitaxial growth at a rate of the order of microns per hour. A simple example is trimethyl-Gallium (TMGa) and arsine (AsH$_3$) in a H$_2$ carrier react (Ga(CH$_3$)$_3$ + AsH$_3$ → GaAs + 3 CH$_4$) forming epitaxial GaAs monolayers.

In principle any number of sources can be attached to a growth reactor. Switching between these enables the layer-by-layer growth of heterogeneous semiconductors within limits set by material properties of strain and reactivity and materials specific residual background levels that may accumulate in the reactor.

Further techniques, such as low pressure LPCVD, are varations on the same theme, each with its strengths and weaknesses. Overall however MOCVD is much used due to its relatively low cost and the monolayer control achievable in the best conditions.

The second higher cost growth method is molecular beam epitaxy (MBE). In this ultra high vacuum technique ultrapure precursor solids are placed in radiatively heated graphite Knudsen cells attached to the growth chamber containing the substrate. Opening shutters on the cells allows a molecular beam to be emitted from the cell at a temperature controlled rate. This beam impinges on the temperature controlled substrate stage, which may be angled to adjust growth modes and conditions, and rotated to optimise growth uniformity.

A range of Knudsen cells are usually attached to a MBE reactor in order to deposit heterogeneous structures on a single substrate, with the same limitations due to geometry and materials properties. Here variants again exist, for example gas-source GSMBE or metalorganic MOMBE.

X.2.3 Heterogeneous growth

The layer deposition methods we have outlined above allow excellent two dimensional control of different materials, but there are fairly tight limitations on the heterostructures that can be grown. The first, which we will not mention in detail, is that certain species with high sticking coefficients for example have an unfortunate tendency to haunt growth chambers for example by dynamically adsorbing and desorbing from surfaces in the growth chamber. This can seriously contaminate subsequent layers and must be avoided by growth chamber purges which significantly increase machine down time and deposition cost. Furthermore, ideal growth conditions differ for different materials. In particular, different growth temperatures are routinely needed for different materials but must be carefully optimised to take account of different thermal coefficients of expansion, of solubility, and therefore of elemental species migration. The greatest difficulty in this class is generally dopant diffusion, as in the well known case of highly mobile Zn diffusion in AlGaAs/GaAs.

Finally, an all-important heterogeneous growth consideration is the lattice constant. Sequentially growing layers with different lattice constants in the same stack gives rise to strain. The total strain energy increases with the layers thickness deposited. Above a limit known as the Matthews-Blakeslee[22] critical thickness, the cumulative strain energy density at the heteroface becomes greater than the bond energy and the total system energy releases strain potential energy by breaking bonds in the interface region. This is strain relaxation which generates dislocations that seriously compromise cell performance and even structural integrity. The Matthews-Blakeslee limit is a function of materials parameters , the lattice constant and elasticity tensors. For example no more than approximately 351Å or so of $In_{0.01}Ga_{0.99}As$, that is about 60 monolayers, can be grown on GaAs before the limit is exceeded and misfit dislocations are generated.

Solutions to the lattice misfit problem have been implemented however. One is strain compensation in multilayer structures with alternating compressive and tensile strain layers. Another is to allow the layer to relax. It is seen that after a sufficient further layer thickness, the dislocation density can return to reasonably low levels, sufficient for some device applications This is known as the virtual substrate or relaxed buffer technique by metamorphic growth. Finally, a variation on this is the graded buffer growth technique[23] which has been used with some success in multijunction solar cells. With this method the composition is varied in incremental steps, restricting the dislocations in each case and again producing an effective virtual substrate.

X.3 Design concepts

Improving design starts with understanding losses. In solar cells under illumination these include extrinsic losses such as reflection or external resistance, and intrinsic, such as optical and electrical transport losses, which we will consider next.

The optical losses result from poor light-matter interaction and first of all, inefficient light absorption. The transport losses can be described under the umbrella of finite carrier lifetimes and corresponding recombination loss via a range of channels.

Understanding and addressing these loss issues can be achieved most reliably via numerical modelling[24-26] which can deliver exact solutions to analytically intractable problems. These are obtained at the expense, however, of some physical understanding , though this can be recovered by sweeping large parameter spaces. These numerical methods are usually required for complex materials and structures.

Understanding can also be gained via analytical methods by applying approximations resulting from exploring physical processes in limiting cases. In this case precisely the reverse is true, in that greater understanding is achieved at the expense of physical accuracy, which may, however, be recovered by refining the theoretical picture. Furthermore, this approach is well suited to crystalline materials and structures symetrical enough to lend themselves to analytical methods, as is the case with many III-V designs.

The following sections follow the second route, and apply analytical models to a range of scenarios. We first develop a picture of the dominant sources of efficiency loss to be addresssed, and investigate some examples of solutions that may address these losses.

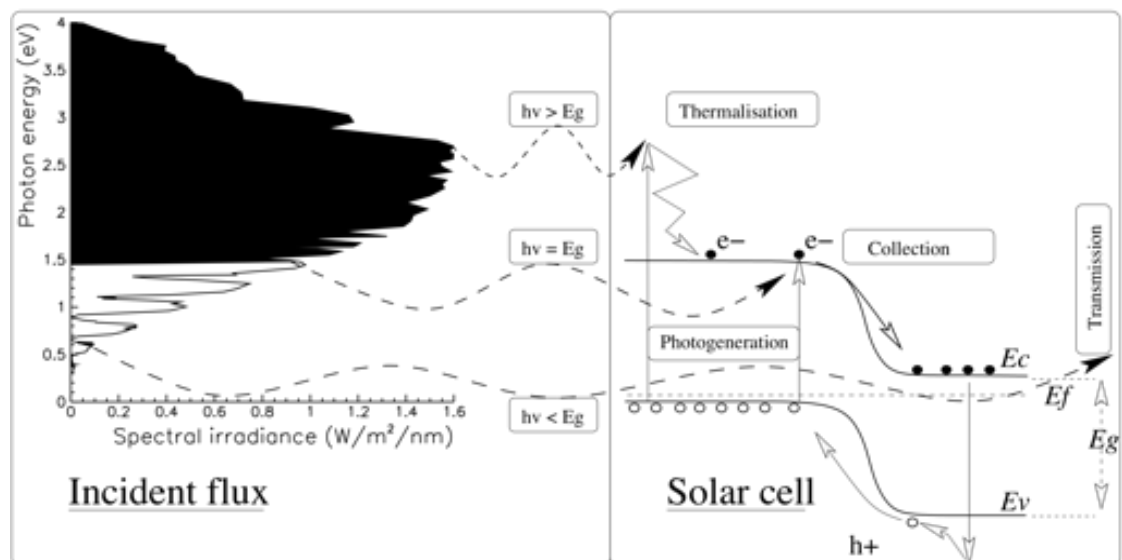

Figure 2  Illustration of losses with respect to the AM1.5G spectrum in a GaAs cell, showing transparence loss for photons with less energy than Eg, and thermalisation loss for electrons and holes absorbing photons with energy greater than Eg.

X.3.1 Light and heat

We now examine the fundamental losses of light absorption and heat dissipation in order to define the basic concepts of solar cell design, and the answers that III-V materials may bring to the issue. Of the first category of optical losses mentioned earlier, we start with the transparency of the cell to photons with energies below its bandgap. This obvious and important fact in cell design is illustrated in figure 2 showing that fraction of the incident AM1.5G spectrum with energy below gap that is transmitted through a cell. The resulting transparency loss as a function of bandgap is illustrated in figure 3. This loss is small for low gap materials as expected, and rises with increasing bandgap. InP, for example, is subject to a 26.6% transparency loss.

The next fundamental loss is thermalisation (fig. 2) whereby carriers photo-excited with energies greater than the bandgap $E_g$ rapidly thermalise, mainly via collisions with the lattice, establishing a steady state minority carrier population with a quasi-Fermi level (QFL) near the band-edge. The photo-generated carriers are harvested with a fixed energy close to that of the lowest energy photons absorbed, wasting the remainder largely as heat. The resulting loss again is shown in figure 3 and this time shows an unsurprising high thermalisation loss for low bandgaps. For InP, again, the loss is a 26.7%, nearly identical to the transparency loss.. This symmetry is consistent with the fact that InP is close to the optimum bandgap as we will see with more exact methods. Together, the transparency and thermalisation loss mechanisms lead to a total maximum efficiency of 46.7% as shown on figure 3 by the solid line combining both loss mechanisms.

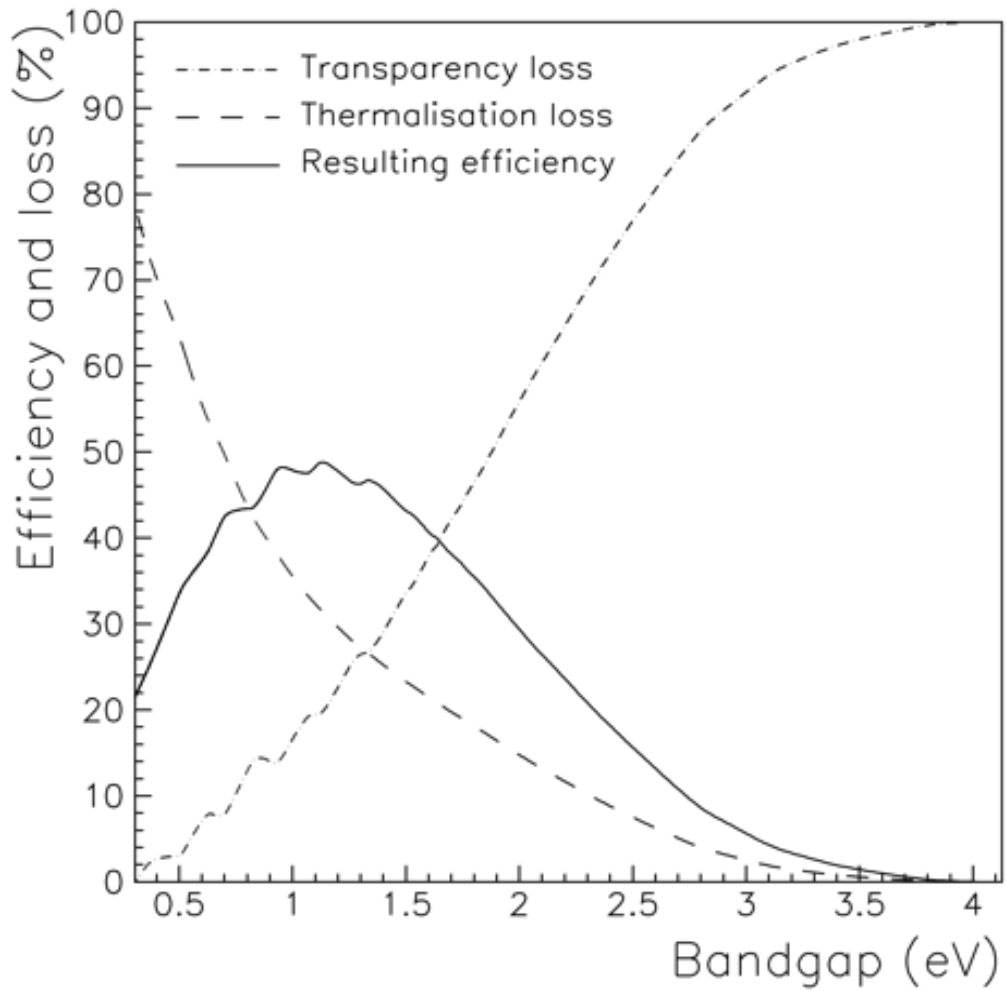

Figure 3 Effects of thermal and transmission losses on single junction solar cell efficiency as a function of cell bandgap considering no other losses.

For a an ideal GaAs cell with no further losses, integrating the potential power as described shows that the total maximum efficiency with only thermal and transmission losses is 45.1%. This limit is set by incident energy losses of 24.8% by thermalisation, and 30.1% through transparency, which is consistent with the slightly higher bandgap of GaAs compared to InP.

Finally, the best case cell efficiency from this analysis is to be 48.8% for a bandgap of 1.13eV, surprisingly close to silicon.

Having set out the basic mechanisms illustrating the trade-off between greater absorption and greater thermal loss, we now develop a more accurate picture of efficiency limiting mechanisms in solar cells, in order to address the resulting issues.

X.3.2 Charge neutral layers

The first transport loss is the well known Shockley[27] injection current in the dark, whereby majority carrier electrons and holes diffuse from an *n* or *p* region across the built in potential or junction bias under a concentration gradient and against the junction potential. They diffuse into a charge neutral region where they are minority carriers, and therefore recombine, giving rise to a net current. The diffusion or injection rate is a function of how long the diffusing carriers remain as minority carriers before recombining – the faster they recombine, the faster they are replaced, thereby increasing the injection current. This is characterised by hole minority carrier lifetimes $\tau_n$ in the *n* doped charge-neutral region and likewise $\tau_p$ for electrons in the *p* doped charge-neutral region, or, via the Einstein relations $L_n=[\tau_n D_n]^{1/2}$ and $L_p=[\tau_p D_p]^{1/2}$ in terms of electron and hole diffusion constants $D_n$, $D_p$ respectively, and diffusion lengths $L_n$, $L_p$ across charge neutral widths $x_p$ and $x_n$. The complete expression[28] for the dark current density at bias *V* can be expressed as

$$J_S(V) = q\left(e^{\frac{qV}{K_BT}} - 1\right)\left[\frac{n_{ip}^2}{N_A}\frac{D_n}{L_n}\left(\frac{\frac{S_nL_n}{D_n}\cosh\frac{x_p}{L_n} + \sinh\frac{x_p}{L_n}}{\frac{S_nL_n}{D_n}\sinh\frac{x_p}{L_n} + \cosh\frac{x_p}{L_n}}\right) + \frac{n_{in}^2}{N_D}\frac{D_p}{L_p}\left(\frac{\frac{S_pL_p}{D_p}\cosh\frac{x_n}{L_p} + \sinh\frac{x_n}{L_p}}{\frac{S_pL_p}{D_p}\sinh\frac{x_n}{L_p} + \cosh\frac{x_n}{L_p}}\right)\right] \quad (1)$$

where $n_{ip}$ is the intrinsic carrier concentration in the *p* layer doped at a level $N_A$, of surface recombination velocity $S_n$, and corresponding parameters $n_{in}$ and $N_D$ in the *n* doped later with its recombination velocity $S_p$.

The Shockley injection formalism, operating under the impetus of a concentration gradient, does not differentiate between bulk transport recombination mechanisms (whether radiative or nonradiative). The lifetimes follow an inverse sum law of contributions from a range of recombination mechanisms, the most important of which are radiative transitions across the gap, and non radiative recombination with phonon emission. The latter usually dominates in charge neutral layers as a consequence of doping as we will see further in this discussion.

We will also see that although the Shockley injection does not discrimate between radiative and non radiative processes, explicitly modelling the upper limit of the radiative recombination can enable us to define an explicitly non-radiative Shockley injection level.

The photocurrent from these layers can be evaluated using standard one dimensional analytical methods in the depletion approximation.[29] We repeat them briefly here to show the complementarity of light and dark solutions and increased model reliability that results, and the understanding that this yields.

Neglecting possible optical reflections at the back contact, The generation rate *G* at position *x* is determined by the Beer-Lambert law relating incident flux *F*, reflectivity *R* and absorption coefficient α as

$$G(x,\lambda) = F(1-R)\alpha e^{-\alpha x} \quad (2)$$

The resulting photocurrent collected from thecharge neutral fraction of the $p$ layer can be evaluated from the excess minority carrier concentration $\Delta n_P$, relative to equilibrium in the dark. This is determined, in the absence of an elecric field term in the charge neutral layers, from the balance of generation and recombination, expressed in terms of $\Delta n_P$ in the $p$ layer as

$$\frac{d^2 \Delta n_p}{dx^2} - \frac{\Delta n_p}{L_n^2} = -\frac{G(x)}{D_n} \qquad (3)$$

which can be solved with appropriate boundary conditions given by surface recombination and the depletion approximation.[3] A similar expression for $\Delta p_n$ yields excess minority carrier concentration in the charge neutral $n$ layer. The photocurrent from both charge neutral layers is given by the gradient, with the appropriate sign, taken at the $p$ and $n$ layer depletion edges, of the minority carrier concentrations. The sum of these two photocurrents defines the charge neutral layer contribution to the total solar cell photocurrent $J_{PH}$.

X.3.3 Space charge region

The space charge region (SCR) non radiative recombination dark current can be expressed in terms of hole and electron diffusion profiles extending across it. This is the Shockley-Read-Hall (SRH) formalism[30] which may be expressed analytically in terms of carrier densities $n$ and $p$ as a function of depletion layer extending from $x_1$ to $x_2$ giving the following SRH recombination current density

$$J_{SRH}(V) = q \int_{x_1}^{x_2} \left( \frac{p(x)n(x) - n_i^2}{\tau_n(p(x) + p_t) + \tau_p(n(x) + n_t)} \right) dx \qquad (4)$$

This current describes the non radiative recombination considering only mid-gap trap levels (the most efficient for recombination) and a space-charge layer with trapped electron and hole densities $n_t$ and $p_t$, and electron and hole non radiative lifetimes $\tau_n$ and $\tau_p$, respectively.

In the space charge region under illumination, the injected majority profiles are perturbed by a small population of free carriers collected at the depletion edges, and of free carriers photogenerated in the SCR. The transport of these excess free carriers is dominated by drift, and lifetimes much greater than the short transit time across the depletion layer. For this reason, the photocurrent contribution from the space charge region is assumed equal to the integral of the generation rate over that region. The sum of this SCR photocurrent and the charge-neutral contributions defines the total photocurrent $J_{PH}$.

X.3.4 Radiative losses

The last loss mechanism we consider is the radiative loss which applies in some measure to both charge neutral and SCR regions. This is the loss that is a direct consequence of absorption, and sets the fundamental limit on the efficiency of solar cells[31].

The generalised Planck equation expresses light emitted by a grey-body[32] as a function of absorption, geometry, and chemical potential or QFL separation of recombining species. It defines[33] the total current density $J_{RAD}$ corresponding to the emitted luminescent flux at bias $V$ from a radiative emitter as an integral over the photon energy $E$ and surface $S$ as

$$J_{RAD}(V) = q\int_0^\infty \left( \frac{2n^2}{h^3 c^2} \left( \frac{E^2}{e^{(E-q\Delta\phi)/kT} - 1} \right) \int_S \alpha(E,\theta,s) dS \right) dE \qquad (5)$$

where $n$ is the refractive index of the grey-body, $\Delta\phi$ is the quasi-Fermi level separation (the difference between hole and electron QFLs), and the other symbols have their usual meanings. The absorptivity $\alpha(E,\theta,s)$ is the line integral over position through the different layers of the cell along the optical path of radiation at angle $\theta$ with the normal exiting or entering surface $S$, the total emitting surface in three dimensions.[32] Therefore, $J_{RAD}$ is minimised by reducing $S$, for example by coating the cell with reflective materials except on the absorbing fraction of the cell's surface facing the sun. This incidentally increases light traping and is closely related to photon recycling concepts.

The spatial variation of the quasi-Fermi level separation is a function of material quality: In the high mobility SCR, it is essentially equal to the applied bias, and constant. In defective (for example heavily doped) charge neutral layers, injected carriers have short lifetimes. Their density therefore decreases exponentially away from the SCR edge, which is equivalent to a QFL separation tending to zero. In thin, high purity charge neutral material, however, lifetimes are long and the diffusion length may be significantly greater than the charge-neutral layer thickness. In this case, it is reasonable to assume, as Araujo and Martí,[31] the upper limiting case of a constant quasi-Fermi level separation equal to the applied bias across the entire device and a consequently higher radiative recombination current.

These points are important in determining how close to the radiative limit cells are operating. They do so by placing a maximum possible upper limit on the radiative character of charge-neutral layers in the case that these layers have lossless transport. This limit can be expressed by evaluating equation 5 with an absorptivity path integral $\alpha(E,\theta,s)$ across the charge neutral layers. In this way, the QFL separation and absorption of different layers and their position determines their contribution to the total luminescence. In this way we can express the upper limit $J_{RAD}^{cn}$ on charge neutral luminescence and corresponding recombination current.

The minority carrier transport is fixed by the carrier continuity equation 6, the solution of which yields the charge-neutral contribution to the photocurrent $J_{PH}$. Knowing the total Shockley injection $J_S$ and the *upper* limit on its radiative fraction $J_{RAD}^{cn}$ we can define the *lower* limit of the non-radiative fraction of the Shockley injection $J_S^{cn}$ as the remainder. That is, the lower limit of the non-radiative fraction of the Shockley injection is $J_S^{cn} = J_S - J_{RAD}^{cn}$.

To summarise these points, the photocurrent and Shockley formalisms complement the radiative limit. The combination of these three formalisms enables us to formulate an explicitly non-radiative modification of the generic Shockley injection level and to define a radiative recombination fraction as a function of bias. This is defined as the radiative fraction of the total recombination as follows:

$$\eta_{RAD}(V) = \frac{J_{RAD}}{J_{SRH} + J_{RAD} + J_S^{cn}} \qquad (6)$$

where the definition relies on the explicitly non-radiative $J_S^{cn}$ in the subscript to avoid double-accounting for the charge neutral radiative recombination current which is already included in $J_{RAD}$.

This important point allows us to clarify an issue whereby dark current measurements and their diode ideality factors cannot differentiate between non-radiative and radiative limit recombination regimes in the dark. These regimes, and hence the radiative recombination fraction defined above, remain relevant in the light, via the superposition principle that we will come to below.

The significance of these remarks on explicit radiative recombination fraction is that, as we mentioned earlier, the radiative limit is the most efficient operating regime of solar cells, in which the only loss is the re-emission of light at an intensity partly determined by the absorption coefficient as shown in the Planck non-black-body equation. It is instructive to look at the resulting maximum conversion efficiency shown in figure 4. The ideal bandgap of 1.35eV for the AM1.5G spectrum is remarkably close to InP, which is wholly consistent with the approximate analysis of thermal and transparency losses seen earlier. A similar analysis can be applied to multiple junction designs as we will see subsequently.

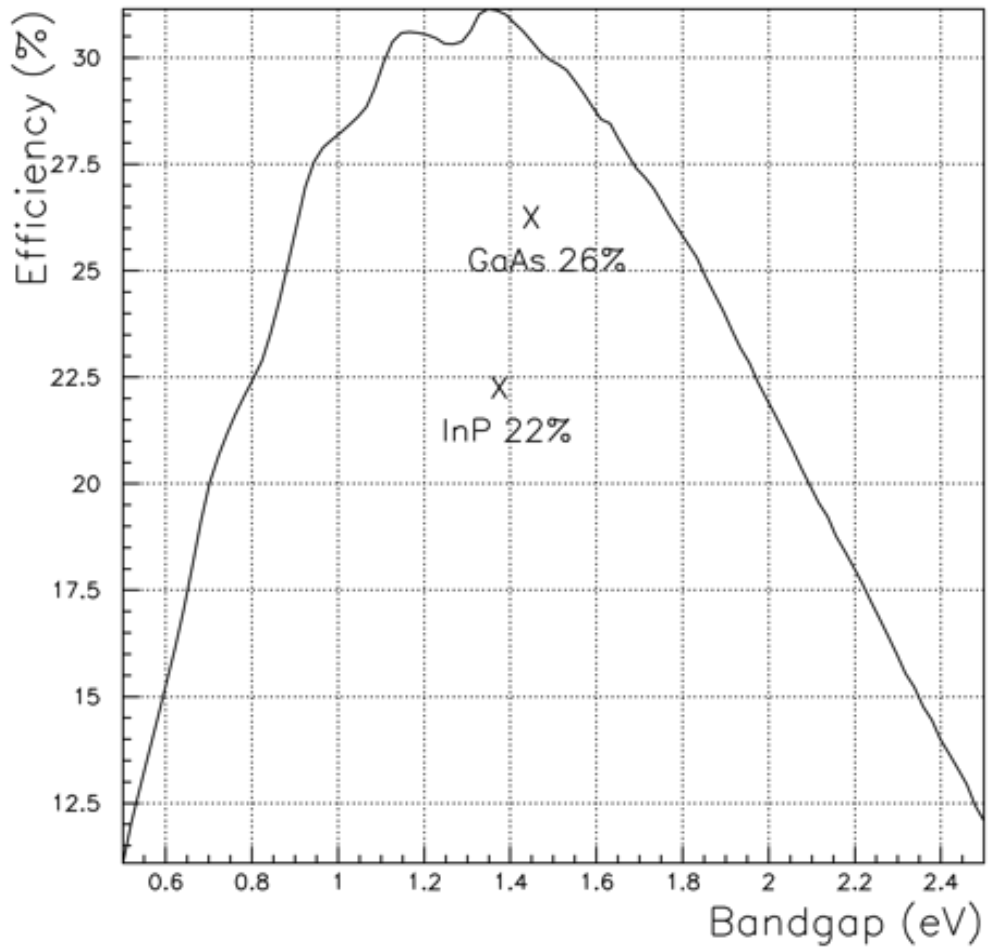

Figure 4 Single junction solar cell ideal conversion efficiency in the radiative limit as a function of bandgap showing best single junction results to date for two key III-V semiconductors. Both have bandgaps close to the optimum of 1.35eV with a potential conversion efficiency of 31.1% for an ideal cell with only radiative recombination losses.

X.3.5 Resulting analytical model

The sum of contributions from charge neutral $p$, and $n$ zones, and space charge regions gives the total photocurrent density $J_{PH}$. This defines the external quantum efficiency including reflection loss (QE) as ratio of collected carriers to number of incident photons at a given wavelength, that is, the probability that a photon incident on the solar cell gives rise to a charge carrier collected at the cell terminal.

Finally, the light current density under applied bias, assuming superposition of light and dark currents is given by

$$J_L(V) = J_{PH} - (J_S + J_{SRH} + J_{RAD}) \quad (7)$$

where we use the photovoltaic sign convention of positive photocurrent.

This light IV enables us in the standard manner[29] to evaluate solar cell figures of merit such as the short circuit current $J_{SC} = J_L(0)$, the maximum power point $V_{MP}$, fill factor FF. Effects of parasitic resistance are included when modelling real data in the usual manner, that is, a series resistance defining a junction bias, and a parallel resistance and associated shunt current reducing the photocurrent.

To put this model in context of other work, it is, firstly, equivalent with the classic Henry model for single to multijunction cells[34] giving 31% efficiency for a single junction in the radiative limit, as seen above (fig. 4). It also agrees with further development concluding with Araujo and Martí[31] and references therein. These authors consider an optimum radiatively efficient design with unit QE, no non-radiative losses, and emission losses restricted to the solid angle subtended by the sun and calculate a limiting efficiency of 40.7% for a single unit gap cell. The method used here is in good agreement, giving 40.1% in the same conditions.

A further contextual issue is that of real data and the capacity of the model to fit it. With regards to this, a result of this modelling approach is the additional constraints placed on variable parameters by the light and dark mechanisms. For example, the minority carrier transport properties are constrained by their specification of photocurrent collection, and also, with a symmetry that reflects the minority carrier origin of both effects, are constrained by the Shockley injection current. This results in fewer free parameters and a more exact understanding of efficiency limiting processes.

The only remaining free parameters are parallel and series resistive losses, and non-radiative lifetimes for electrons and holes in the space charge region. Both of these however are adequately constrained by the dark current data and in their respective bias ranges constitute single parameter fits.

The examples given so far have considered only AM1.5G solar spectra. The same principles hold for other spectra which we introduce here, such as the AM1.5 direct AOD and spatial AM0 spectra, together with light concentration frequently used in the field.

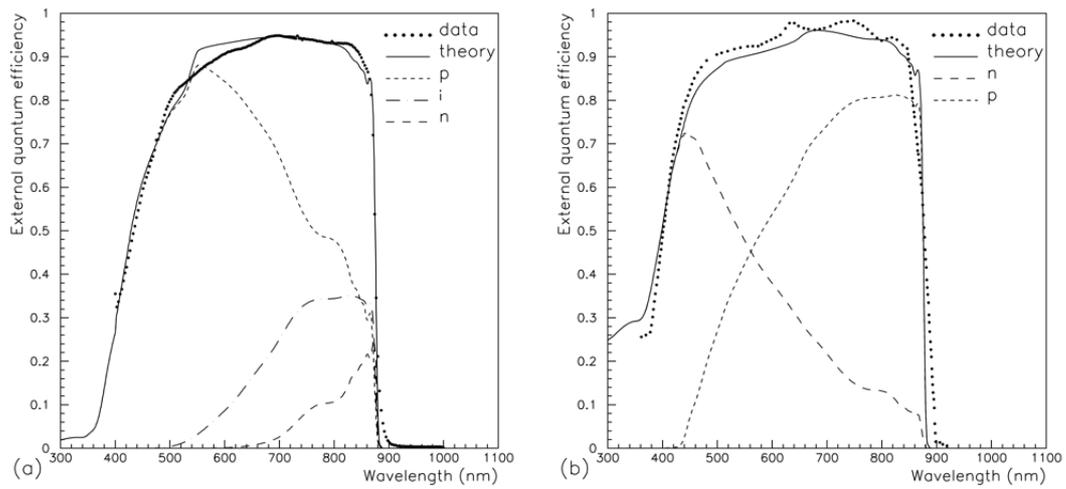

Figure 5 Spectral response of GaAs cells, showing a 20% efficient *pin* structure (a) and a record 25% efficient *pn* cell (b).

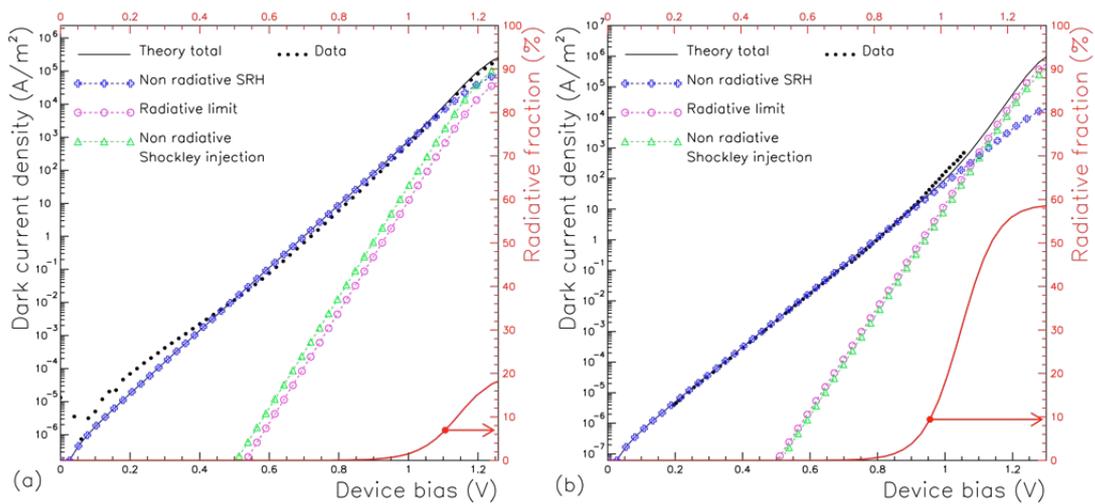

Figure 6 Dark current of a 20% efficient *pin* GaAs cell (a) and the record 25% efficient *pn* GaAs cell (b). The higher conversion efficiency for the *pn* cell is consistent with its higher radiative recombination fraction: at high bias approaching flat band, these recombination fractions reach 58% and 18% respectively, as indicated on the right axes of both figures.

X.3.6 Single junction analyses

To illustrate the concepts developed above, we look at two example of single junction solar cells. The first is a well characterised non-ideal 20% efficient *pin* structure (extrapolated to 5% shading) comprising a nominally undoped intrinsic *i* layer between emitter and base, while the second is a record 25% GaAs *np* solar cell[35] with less available data but showing a slightly different and superior operating regime, and also with 5% shading reported.

Figure 5 shows the spectral response data and model for both cells. The first notable difference is the significant intrinsic region contribution in the 20% *pin* cell as would be expected, and the good fit resulting from the use of the measured reflectivity (not shown) in the modelling.

The 25% *np* cell shows a slightly inferior fit, resulting from the need to calculate the front reflectivity for a dual layer $MgF_2$/ZnS anti reflection coating as described in the Kurtz reference[35] (using 120nm $MgF_2$ and 65nm ZnS thickness rather than the inconsistent 6.5nm quoted in the reference). It shows a negligible SCR contribution and a significantly higher short wavelength response than the 20% cell, despite the less promising *np* geometry for which the short wavelength QE is dominated by less efficient hole minority carrier collection in the n-type emitter layer. This is related to the main novelty of this cell, which is the use of a 30nm GaInP window on a thin 0.1µm n-type emitter. The novel window is responsible for very low emitter-window recombination velocity allowing a thin emitter without excessive speading resistance and high collection efficiency in this thin n-type layer.

Figure 6 shows the complementary modelling in the dark, using the transport parameters consistent with the QE modelling. The left and right axes show respectively the dark current contributions and resulting radiative efficiency.

The 20% cell never reaches radiative dominance, the radiative share of recombination reaching about 18% as the cell approaches flat band and the effects of series resistance start to appear. More importantly, the radiative fraction at the maximum power point $V_{MP}$ under one sun illumination is just 0.1% showing overwhelmingly non-radiative dominance in this cell.

Although dark IV data for the 25% efficient cell is not available to similarly high bias, the modelling of the lower efficiency cell strengthens the analysis of what data is available. Additionally, the fit is not as exact, which is attributable in part to less precise knowledge of reported cell geometry and in particular cell grid coverage reported as approximately 5%.[35] Bearing these issues in mind, the overall agreement, summarised in table 1, is nevertheless close.

Comparing the two cells, we note that the radiative current density is comparable if slightly higher in the *pn* as might be expected in the light of the differences in geometry and shading which are such as to have little bearing on the net rate.

The non radiative SRH rate is however much greater in the *pin* structure, despite the electron and hole SCR lifetime of 10ns as opposed to just 2ns in the 25% *pn* cell. This apparent contradiction between the longer lifetime in the less radiatively efficient *pin* cell and the lower non radiative recombination rate in the *pn* cell has two causes: The most important is the obvious longer lifetime in the undoped *i* layer of one structure which means less dopants or, equivalently, a lower defect density, and a longer lifetime as shown by the modelling. The lower non radiative injection current in the more efficient *pn*, on the other hand, is explained in part by the superior performance of the n-type charge-neutral layer in the *pn* case as a result of the novel window layer at the time of publication.

The overall conclusion is that the modelling is consistent with available light and dark data and suggests that the 25% record cell is just about radiatively dominated but only at high bias. That is, the explicitly radiative recombination from the SCR and charge neutral layers accounts for 58% of the total as the cell approaches flat band. In addition, series resistance is negligible in this case, reflecting the high quality GaInP of window layer design and consequent high conductivity of the solar cell surface layers with little loss of photogeneration. This represesents the highest radiative efficiency this cell can conceivably attain at the high current levels obtained at high illumination levels under concentration. More practically, and more importantly, is the situation at the maximum power bias $V_{MP}$=0.91V (table 1). At this bias under one sun illumination, the radiative recombination fraction is 4%. This is far higher than the less efficient 20% cell, and yet still overwhelmingly non radiatively dominated.

There emerges a consistent picture of the physical phenomena developed in describing these high purity crystalline solar cells: The dark current and light current modelling consistency leads to constrained modelling which reveals detailed information concerning the operational regime of solar cells.

One conclusion of looking at the radiative fraction in the high bias regime where ideality 1 starts to dominate is that a solar cell with an ideality of 1 may be far from the radiative limit. It may in fact only ever asymptotically approach the radiative limit as doping levels in the charge-neutral layers are decreased, hence reducing the doping related defect density and non radiative recombination rate. In this low doped case, however, the overall cell efficiency drops due to a significant reduction of the built-in potential relative to the cell bandgap. An optimum can be estimated with the analytical methods described. A detailed analysis is beyond the scope of this chapter, but we can say that the optimum is a trade-off between high doping levels and efficient transport. High doping ensures a high junction potential, and lower injection. But shorter neutral layer lifetimes imply both higher injection levels and lower collection efficiency. A proper optimisation in terms of these competing processes ensures high collection efficiency together with a low Shockley injection, consistent with tending asympotically towards the radiative limit.

|         | Jsc (A/m2) | Vmp(V) | Voc (V) | FF (%) | Efficiency (%) |
|---------|------------|--------|---------|--------|----------------|
| Kurtz[32] | 285      | *NA*   | 1.05    | 85.6   | 25.0 ± 0.8     |
| Model   | 278        | 0.91   | 1.05    | 82.7   | 24             |

**Table 1**. Record GaAs cell parameters published by Kurtz[32] for AM1.5G compared with analytical model results.

X.3.7 Conclusions

This analytical overview of solar cell performance has examined the trade-off between thermal and transparency losses, and suggests that reducing these important losses is a promising strategy.

The more detailed analysis of radiative and non radiative losses has shown a more realistic and significantly lower achievable efficiency with a single bandgap. Analysis of an efficient published cell shows an interesting point which is that solar cells with ideality factors tending towards 1 at high bias are not necessarily tending towards a regime dominated by the highest potential efficiency radiative recombination limit.

These two points analysing single junction performance and loss set the stage for designs going beyond the single junction design in the following sections.

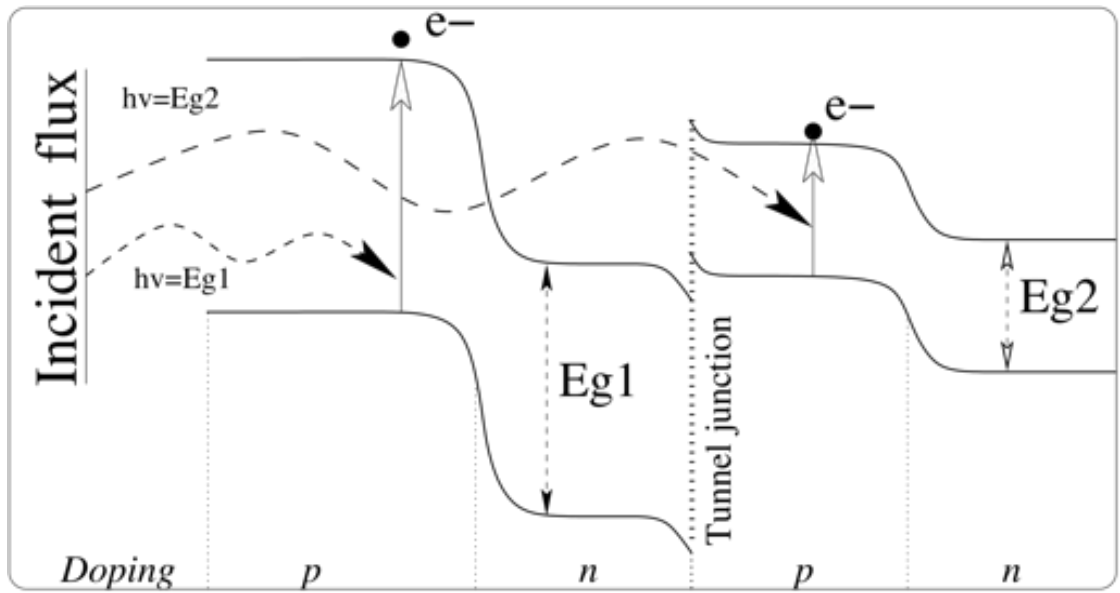

Figure 7 Monolithic tandem multijunction solar cell band diagram solution to thermalisation losses. The series connection of *pn* junctions requires constant series current and a tunnel junction in order to cancel the current blocking np junction formed between *pn* sub-cells. The design raises materials compatibility issues for monolithic cell growth, in particular lattice constants.

X.4 Multijunction solutions

X.4.1 Theoretical limits

In order to reduce the fundamental losses illustrated in figures 2 and 3, we must first absorb all incident photons and yet arrange things such that all these photons are absorbed close to the band-edge. These conflicting requirements could be resolved by reshaping the spectrum either by means of an intermediate filter absorbing all incident photons and re-emitting light in a narrow spectrum or ideally as a monochromatic beam. This method, and its variants up and down conversion,[36][37] needs only a single junction accepting the re-shaped spectrum but predictably suffers from efficiency losses in the spectral conversion.

Another option is spectral splitting,[38] whereby the spectrum is separated into different, ideally monochromatic beams, which are absorbed by solar cells with appropriate bandgaps tuned to the part of the spectrum they are designed to convert to electrical power. This is a multiple cell solution, where, for most applications, the sub-cells would be connected in parallel, or in series with an equal series curent constraint.

This concept of spectral splitting finally leads us to a simpler solution, developing the notion of sub-cells, which is to achieve a similar result by arranging the sub-cells optically in series, each acting as an optical filter to those underneath it. As shown in figure 7, the first cell to see the spectrum converts and filters the high energy photons, and so on through ideally an infinite number of junctions. This is the multijunction solar cell, and is ideally suited to III-V solar cell materials since, as we saw at the start of this chapter, these materials cover the greater part of the solar spectrum.

The multijunction solution raises the problem of how to connect the sub-cells. The mechanically stacked solution is to place them in series optically, and contact them individually in parallel or even completely independently. This has the advantage of allowing the combination of arbitrary materials which may be lattice and current mismatched. However the complexity resulting from the multiple connections and optically efficient stacking means that this technique is limited to concentrator arrays.[1]

Another solution is the monolithic series connected design, illustrated in figure 7 for a tandem cell. This scheme requires a constant series current constraint through all sub-cells, and is therefore limited by the lowest photocurrent contribution. The other cells are forward biased away from their maximum power point until the series current constraint is met, resulting in a loss in output power. It also implies compatible growth in principle, regarding lattice constants and growth methods.

A final design issue is the reverse diode presented by the series connection of subsequent sub-cells, illustrated in figure 7, whereby the *pn* junction sub-cells are inevitably connected by a reverse biased *np* junction acting as a blocking diode. This must be short-circuited by a highly doped tunnel junction[39] allowing majority carriers to flow unimpeded between sub-cells, thus completing the circuit.

A simplified model sufficient for our purposes is given by Demassa[40] and allows calculation of tunnel junction characteristics in terms of bulk materials parameters of the layers defining the tunnel junction, in particular effective masses, permittivity, and doping levels. A brief calculation which we will not describe in detail shows that good tunnel junction materials must posess a high density of states, and must be degenerately doped. Solar cells impose further constraints, which are that the tunnel diode must be as thin as possible, and possess a high bandgap in order to remain optically thin, since any light absorbed in these degenerately doped layers does not contribute to photovoltaic action. A properly designed tunnel junction is ohmic up to a limiting current, and may be treated as a series resistance, together with an associated optical loss, which is the approach used in the modelling presented here.

To conclude this brief mention of ohmic tunnel junctions, we note, referring to figure 1, that AlGaAs and GaInP, lattice matched to GaAs, are good candidates for tunnel junctions because of their high bandgaps. Another material mentioned earlier is AlInP lattice matched to GaAs, and which constitutes another indirect high bandgap material for tunnel junctions and window layers as we will see subsequently.

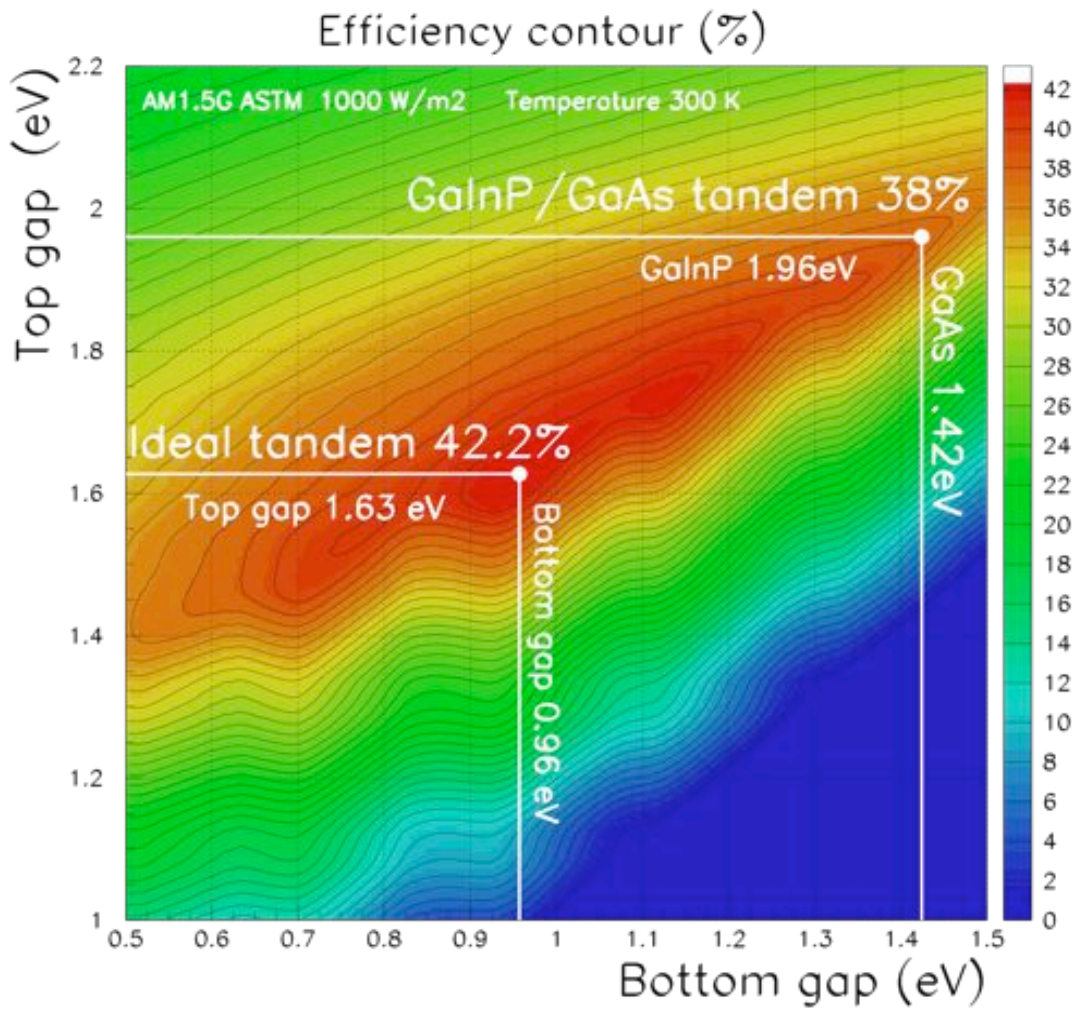

Figure 8 Ideal dual junction solar cell maximum conversion efficiency in the radiative recombination limit as a function of higher and lower junction bandgaps, showing the absolute maximum of 42.2% and the highest conversion efficiency of 38% achievable with a GaInP on GaAs tandem.

X.4.2 Materials limitations

The radiative efficiency limit for an infinite number of sub-cells[31] is about 86%. More practically, we find that a tandem cell with two junctions may reach 42.2% without concentration. This is illustrated by the efficiency contour (figure 8) in terms of upper and lower gaps assuming only radiative losses and therefore a perfect, lossless tunnel junction. Non-ideal bandgap combinations may, however, reach efficiencies close to this. Examining at the contour shows a tandem efficiency that is relatively insensitive to change in bandgap as long as both gaps are varied simultaneously. For example, a tandem with gaps (0.8, 1.6)eV will perform roughly the same as one with gaps (1.19, 1.78)eV, with an efficiency of about 42%, both only slightly lower than the absolute maximum.

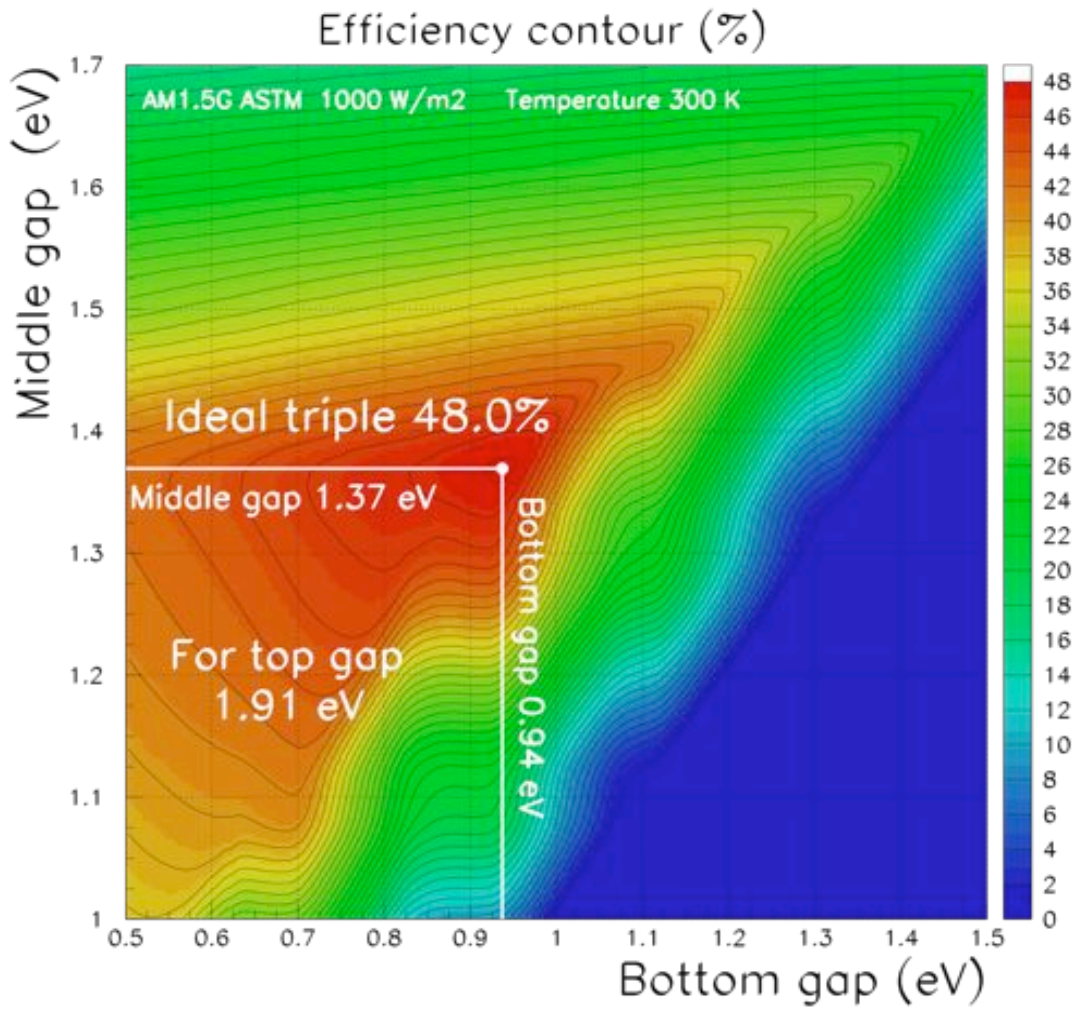

Figure 9 Ideal triple junction solar cell maximum conversion efficiency in the radiative recombination limit showing a section through the triple gap bandgap/efficiency volume, by fixing the highest gap component at its ideal value of 1.91eV. The world record triple at gaps 1.87, 1.40, and 0.67 eV of conversion efficiency 40.7% under a different, concentrated spectrum cannot be shown on this contour given its non-ideal top sub-cell bandgap.

Nevertheless, considering the ideal limit first, the upper and lower bandgaps for the absolute maximum tandem efficiency of 42.2% are 1.63eV and 0.957eV, as shown on the figure 8. For the most promising substrate, GaAs, figure 1 shows two materials with the higher gap (1.63eV) that are lattice matched to GaAs.

The first, $Al_xGa_{1-x}As$, remains direct from x=0 up to $Al_{0.49}Ga_{0.51}As$, a bandgap ranging from 1.424 to about 2eV and including the 1.63eV cell for a composition of approximately x=0.17.

The second possible material is in the $Ga_xIn_{1-x}As_yP_{1-y}$ family which ranges from 1.424eV (GaAs) to 1.9eV (that is, x=0.51, y=0.49) with a continuous range of group III and group V compositions lattice matched to GaAs. This same flexibility raises another important advantage, which is the possibility of lattice matching this quaternary to Ge, and even to Si substrates.

Ironically, this phosphide material has been much studied on InP substrates for telecommunications applications,[7] but has received little attention on GaAs substrates because of the availability of AlGaAs which is historically well established, and easier to grow [6], despite its non ideal minority carrier characteristics. The consequence is that materials knowledge is largely restricted to the lattice matched ternary endpoint $Ga_{0.51}In_{0.49}P$, and that lattice matched quaternary materials are simply expressed by a linear interpolation as $(GaAs)_{1-z}(Ga_{0.51}In_{0.49}P)_z$.

These considerations suggest that this quaternary materials family is worthy of greater attention. However, in the current state of knowledge, the quaternary composition $(GaAs)_{0.8}(Ga_{0.51}In_{0.49}P)_{0.2}$ has the correct direct gap of 1.63eV for our ideal tandem structure, for which we need to identify a lower gap 0.957eV material.

For this lower gap however there is no lattice matched candidate. Using GaAs, the lowest available gap, as the lower gap sub-cell of a tandem yields an ideal efficiency limit of 38%. This dictates an ideal upper sub-cell bandgap of 1.95eV, obtainable with AlGaAs but approaching the indirect transition for this material where the recombination associated with the DX centre corresponding to the L indirect valence band minimum becomes increasingly important.[11] For these reasons, AlGaAs is generally not considered as a candidate for tandem cells and we will not consder it further. The GaAs based tandem, however, remains a viable design with a compatible phosphide material which is nearly ideally matched by GaInP, and has the potential to reach 38% efficiency.

Coming back to the ideal tandem efficiency limit of 42%, the closest material with the correct lower sub-cell bandgap of 0.957eV is $In_{0.43}Ga_{0.57}As$. This compound is lattice mismatched to GaAs substrates, with a critical thickness of just 8nm after which misfit dislocations result in a serious penalty in cell efficiency.

Figure 9 shows another step on the road to the 86% limit consisting of three junctions, showing the result of a numerical optimisation of the three gap system efficiency in the radiative limit. In this case, the efficiency contour is much sharper than for the tandem case. Any deviation from the ideal brings a rapid decrease in efficiency that cannot be corrected to the same degree by adjusting the other two bandgaps.

The ideal sub-cell bandgaps found for the triple multijunction cell are 1.91eV, 1.37eV and 0.94eV, which together give an efficiency of 48%. Referring to figure 1, we find, again, that the most promising materials belong to the GaInAsP family and indeed are significantly closer to simple ternary compounds. The upper gap is well approximated by $Ga_{0.515}In_{0.485}P$ of gap 1.9eV which is lattice matched to GaAs. The middle gap cell at 1.37eV corresponds to $In_{0.05}Ga_{0.95}As$ which has a critical thickness of 70nm on a GaAs substrate. Finally the lowest gap varies little compared with the tandem case, and therefore presents the same problem as in that case: the lack of lattice matched lower gap materials.

This survey of the materials available in III-V semiconductors for ideal tandem and triple junction cells shows that even this wide range of materials requires some additional tricks to circumvent materials issues resulting from mismatched materials. Our brief exploration of available materials leads us to conclude (somewhat counter intuitively given the range of materials available) that the most immediately promising materials for both tandem and triple junction designs are GaAs and GaInP, and an as yet undetermined lower gap material for the triple. The following sections examine some solutions to these issues and the record breaking multijunction cells that have been fabricated as a result.

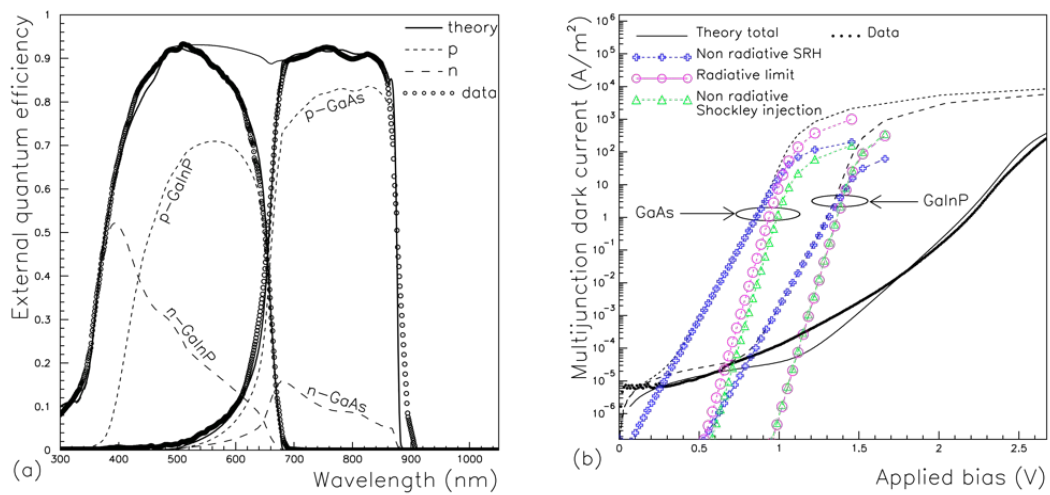

Figure 10 QE data and modelling showing the detail of layer contributions (a) and dark current data and modelling (b) for a high efficiency GaInP/GaAs tandem (Japan energy Corp. [39]).

X.4.3 A tandem junction example

The previous sections have given some understanding of the sources of efficiency loss, of how to moderate them, and of candidates amongst the III-V materials where these ideas may be put into practise.

Among the large body of work on multijunction cells (see for example Andreev[1]), we now present an analysis of a tandem consisting of the GaAs/GaInP combination we mentioned earlier with a theoretical maximum efficiency of 38%. This has been attempted by a number of groups. One of the first achieved over 30% under concentration of 100 to 200 suns in 1994.[41] We will focus on a later result from 1997 by Takamoto *et al.*[42] to whom we are indebted for quantum efficiency and dark current data. The Takamoto paper reports over 30% efficiency under a global unconcentrated spectrum.

The full devices structures are available in Takamoto's paper and we mention only the main points here. The device structure chosen is *n* on *p* with an AlInP window layer. Like the Kurtz single junction cell, the n-type emitter is heavily doped and only 50nm thick, whereas the lightly doped base (hence with good minority carrier transport) is over an order of magnitude thicker at 0.55μm. The tunnel diode comprises of *n* and *p* doped InGaP layers of 30nm each doped $10^{25}m^{-3}$ (Zn) and $0.8\times10^{25}m^{-3}$ (Si) respectively, and sandwiched between higher gap AlInP cladding layers. These are intended, in part, to reduce the common problem of dopant diffusion from the highly doped tunnel junction. The device is completed by a GaAs bottom junction with a InGaP back surface minority carrier reflector.

Figure 10a shows the modelled QE again assuming a double layer $MgF_2/ZnS$ antireflection (AR) coat and showing a good fit overall. The breakdown of different regions clearly shows one strength of the design which is the dominance of photocurrent produced by p-type layers which benefit from better electron minority carrier transport than the n-type layers. In this *n* on *p* geometry, the more efficient *p* layer dominance is achieved by the use of a thin *n* layer and combined with good interfaces between the GaAs *n* layer and the tunnel junction cladding, and the topmost n-type GaInP layer and the AlInP window.

A further important point to note is the significant QE of the bottom GaAs cell above the GaInP bandgap: this means that the top cell is some way from being opaque near its bandgap. This is another design feature of this cell: In order to ensure current continuity in the light of the imperfect band-gap combination imposed by the GaAs as discussed above, the top cell is made thinner in order to ensure current continuity at the expense of a slightly greater thermalisation loss in the GaAs cell.

Figure 10b shows the corresponding dark current fit together with individual subcell and overall tandem dark currents. The tandem current is determined by the sub-cell current-voltage characteristics by adding sub-cell biases at constant current assuming an ohmic tunnel junction. As mentioned earlier, this estimates the Shockley injection from the transport parameters, and calculates the radiative current from the cell geometry and absorption coefficient, both validated by the QE fit. In the case of the dark current there is, in this case, an imperfect fit and signs of systematic error at low bias, which cannot be elucidated further without separate dark IV curves for both subcells. At higher bias however, agreeement is sufficient to indicate that the higher gap GaInP sub-cell is approximately 40% radiative, and the bottom GaAs sub-cell approximately 65% radiative, comparable if slightly better than the Kurtz single junction cell reviewed previously.

The model results are compared with the published data in table 2. In this case, the model slightly under-estimates experiment, due in part to under-estimating the short circuit current density by about 2%, but more importantly because of over-estimating the dark current at high bias by a factor of up to 2 in the region of the $V_{OC}$ at 2.5V. This can be seen in the under-estimate in $V_{OC}$ in particular.

Concerning the dark current over-estimate, the radiative current can only be over-estimated if there is a net reduction of the luminescence. The luminescence from the front surface is small, due to the critical angle for total internal reflection at the front surface of 18% for this design. Most of the luminescence (of the order of 90%) is therefore lost to the substrate where it is effectively absorbed. The only scenario for a significant over-estimation of the radiative current is therefore the presence of a reflective rear surface. This is ruled out by the good quality of the QE fit near the band-edge with the nominal structure (figure 10a), which has no rear reflector.

The only remaining possibility is that the minority carrier transport properties in the charge-neutral layers are better than standard values tabulated as a function of composition in the literature, as used by the model.

However, despite these issues we can conclude that the tandem solar cell operates in a regime which is consistent, and slightly better than, the single junction GaAs cells. That is, the GaAs sub-cell 12% radiative fraction at the maximum power point reported in table 3 is greater than the corresponding 4% reported above in the "Single junction analyses" section for earlier single junction values, which is consistent with marginally superior $V_{MP}$, $V_{OC}$, and $FF$ (cf. table 1). The GaInP sub-cell is comparable if slightly better at 13% radiative fraction at its operating bias $V_{MP}$.

To put this performance in context, this 30% record tandem operates at an impressive 78% of the 38% ideal radiative limit for this bandgap combination, and at 71% of the 42% ideal radiative limit with no materials restrictions.

|  | Jsc (A/m2) | Voc (V) | FF (%) | Efficiency (%) |
|---|---|---|---|---|
| JEC[39] | 142.5 | 2.49 | 85.6 | 30.3 |
| Model | 139.5 | 2.32 | 87.0 | 29.4 |

**Table 2**.Tandem cell published parameters for AM1.5G compared with analytical model results, showing reasonable agreement but with an under-estimated Voc due to an over-estimated dark current.

| Tandem subcell | Vmp (V) | Radiative fraction $\eta_{RAD}(V_{mp})$ (%) |
|---|---|---|
| GaInP | 1.34 | 13 |
| GaAs | 0.926 | 12 |

**Table 3**.Modelled sub-cell radiative recombination fraction at respective maximum power points showing non negligible radiative recombination levels.

| Spectrum | Top gap | Mid gap | Low gap | Jsc (A/m$^{-2}$) | Voc (V) | Eff. (%) |
|---|---|---|---|---|---|---|
| AM1.5G (1000 W/m$^{-2}$) | 1.91 eV | 1.37 eV | 0.94 eV | 167.0 | 3.20 | 47.9 |
| AOD (913 W/m$^{-2}$) | 1.86 eV | 1.34 eV | 0.93 eV | 157.3 | 3.09 | 47.6 |
| AM0 (1354 W/m$^{-2}$) | 1.84 eV | 1.21 eV | 0.77 eV | 235.2 | 2.84 | 44.0 |

**Table 4.** Ideal triple cells gaps and efficiencies for standard global, direct, and near-earth space spectra.

X.4.4 Record efficiency triple junction

The examples discussed in previous sections are in terms of a AM1.5G global spectrum without concentration. The more complex multijunction cells (and, increasingly, single junction cells) are usually reported in terms of the AM0 spectrum just outside the earth's atmosphere, or the terrestrial direct beam AOD spectrum[43] at

AM1.5. Table 4 gives a summary of ideal triple junction characteristics calculated in the radiative limit for these three spectra without concentration. The AM0 spectrum gives the lowest conversion efficiency despite the highest power since this is the broadest spectrum, and as such an ideal cell loses more power below its bandgap in the infrared.

The materials limitations are more stringent than in the tandem case. This is first because the triple junction efficiency is more sensitive to variations in bandgap, and because the ideal bandgaps are further from those of available lattice matched materials.

In this context, these designs and their higher efficiencies have led to the development of lattice matched and heterogeneous growth III-V cells on Ge substrates. An example consisting of GaInP, GaInAs and Ge substrate subcells is provided by King *et al.*[44] where the reference includes the detailed cell structure together with lattice matched and mismatched cells.

The lattice matched material looks at the optimum material quality option, whereas the lattice mismatched, metamorphic option is intended to approach the ideal subcell bandgaps more closely. In addition, this paper covers a further interesting degree of freedom that we have mentioned above, which is the use of group III sublattice disorder to control the bandgap, and thereby the current matching in the triple junction. The band structure is not explicitly stated by King *et al.* but is approximately (1.87, 1.40, 0.67) eV for lattice matched GaInP, $Ga_{0.99}In_{0.01}As$, and Ge subcells.

In a little more detail, the optimal bandgaps in the ideal limit for a Ge substrate sub-cell are 1.88eV for the top GaInP subcell, and 1.33eV GaInAs middle gap sub-cell for a 0.67eV Ge subcell and substrate. The ideal one sun AM1.5G efficiency for this structure is 45.5%.

The GaInP may be engineered to match this with judicious use of composition and ordering mentioned above. The ideal $Ga_{0.955}In_{0.055}As$ however cannot, but the tolerable critical thickness or nearly 3μm for this layer is the reason the metamorphic route is investigated in the King reference.[44]

The net difference between the two cases being relatively minor, however, with similar performance within margin of error. In this discussion, therefore, we limit the analysis to the lattice matched case as a direct progression from the previous single junction and tandem cases examined.

The main interest from the point of view of this discussion is the investigation of ideality 1 and 2 mechanisms reported in the paper for both lattice matched and mismatched triple structures. The method used by King *et al.* is the probing of $J_{SC}$ and $V_{OC}$ as a function of cell illumination intensity. Subject to the assumption of the superposition principle mentioned earlier, this yields the dark current, and is provided for the upper GaInP and middle GaInAs subcells but not the low gap Ge subcell.

Concerning earlier discussion of the meaning of ideality 1 regimes, the paper explicitly defines the ideality n=1 regime as the regime where the dark current is dominated by the Shockley injection current and concludes that the sub-cells increasingly approach the ideality 1 regime at high bias consistent with the modelling reported here.

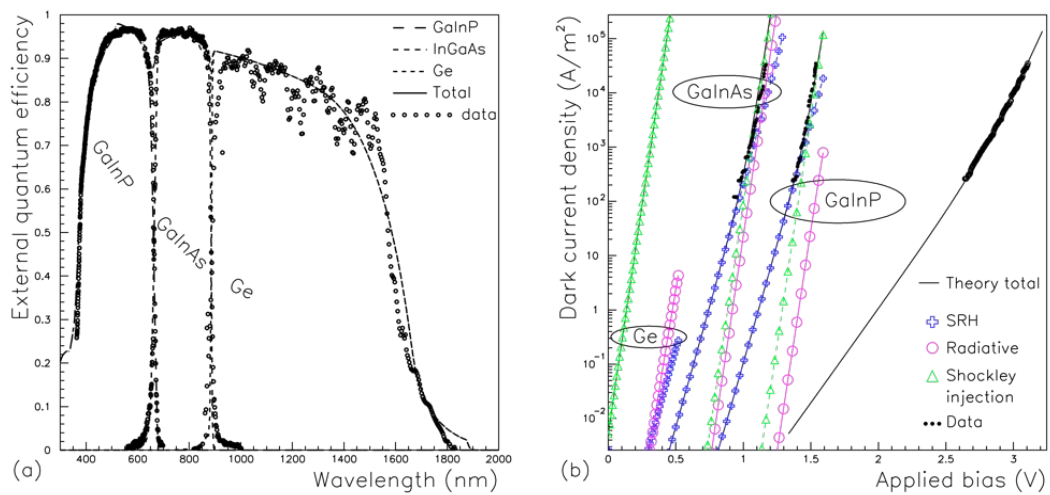

Figure 11 QE data and modelling (a) for the three subcells of a GaInP/GaAs/Ge record triple junction cell [41] and (b) dark current data and modelling for two of the subcells and the combined triple junction response showing the recombination regimes described in the text.

Figure 11 shows the modelled QE for each subcell assuming, as before, a calculated reflectivity consistent with the published data. The modelling uses transport data from the literature,[4-7] validated by the good QE fit. As illustrated by the sub-cell light IV in figure 12, the model shows that, as reported by King,[44] the Ge substrate bandgap is significantly below the optimum and therefore this sub-cell over-produces current. In consequence, is it forced into forwards bias in order to decrease its net current and achieve current parity and continuity with the other subcells, and, therefore, operates at lower efficiency at a bias beyond its maximum power point. The other consequence of a non-ideal lower bandgap is excessive thermalisation in this Ge sub-cell and consequent efficency loss.

The dark current fitting is shown on the right hand of figure 11 for available sub-cell data and the overall experimental tandem response. The lack of Ge data requires the use of transport data from the literature for this material which is nevertheless validated by the good QE fit. The relatively short range of available dark current data for the GaInAs junction, GaInP junction, and overall triple junction device nevertheless covers the transition between the n≈2 and n≈1 idealities, together with the experimental maximum power point $V_{MP}$ and $V_{OC}$ reported.[44] For these three sets of data, the model provides a good fit subject to the proviso of short datasets.

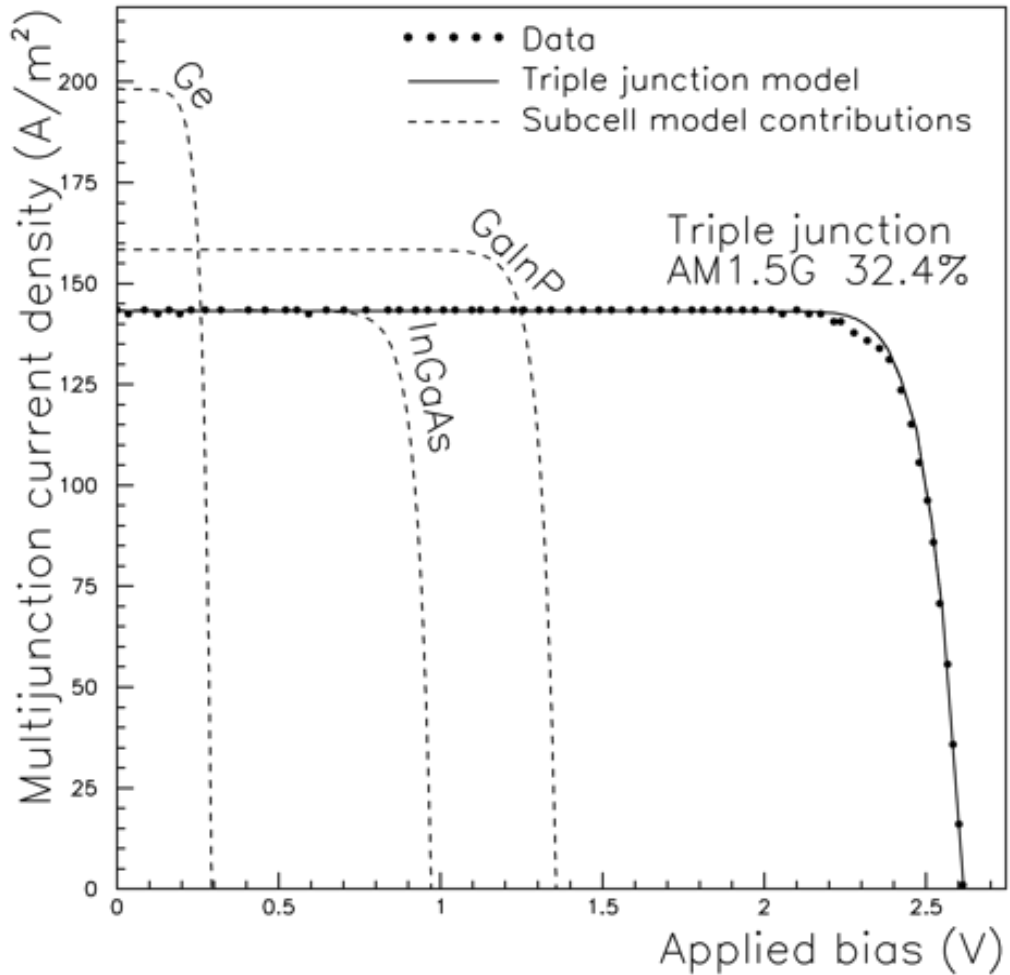

Figure 11 QE data and modelling (a) for the three subcells of a GaInP/GaAs/Ge record triple junction cell [41] and (b) dark current data and modelling for two of the subcells and the combined triple junction response showing the recombination regimes described in the text.

With regards to efficiency, the light current theory and experimental data are shown in figure 12, and corresponding conversion efficiency parameters given in table 5, for aperture area efficiency as reported,[44] which disregards shading losses.

The first conclusion from these results is that the dark current data available is sufficient to analyse the operating regime of this cell at the maximum power point. The combined modelling of the GaInP and GaInAs subcells in figure 12 and table 5 shows that the range of dark current data and modelling covers the illuminated maximum power point bias $V_{MP}$. It also covers the transition from non radiative dominated ideality at low bias to the n=1 ideality marking the Shockley injection regime which may be radiatively dominated.

Taken as a whole, the modelling and experiment for this triple show a close match in the critical transition from non radiative dominated high ideality behaviour at low bias to n=1 behaviour at high bias. In order to explore radiative recombination contribution, table 6 shows the radiative fraction of the total recombination current in each subcell at its maximum power point. This shows that the radiative fraction of the sub-cells is significantly lower than the single junction (4% radiative at $V_{MP}$) and tandem junction records (GaInP and GaInAs operate at 12% and 13% radiative fraction respectively at their $V_{MP}$ points).

The analysis shows quantitatively that this cell, even when favoured by reporting active area efficiency, is significantly further from the ideal limit than the single and tandem junction cells seen earlier. This is to some extent reflected in its proportionally greater difference with the 45.5% limiting efficiency for the sub-cell bandgaps (1.88, 1.33, 0.67 eV) in the same conditions: That is, the cell is operating at 70% of the ideal limit within constraints set by the Ge substrate, and at 66% of the unconstrained radiative limit for an ideal triple junction cell which is the 48% figure we discussed earlier illustrated in figure 9.

|  | Jsc (A/m2) | Voc (V) | Vmp (V) | FF(%) | Efficiency (%) |
|---|---|---|---|---|---|
| King[41] | 143.7 | 2.62 | 2.30 | 85 | 32.0 |
| Model | 143.2 | 2.62 | 2.31 | 86.0 | 32.4 |

**Table 5**. Comparison of modelling and published parameters for the triple juntion[41] lattice matched GaInP/GaInAs/Ge design under the AM1.5G spectrum compared with modelling. Good agreement overall is observed despite the assumptions made regarding the Ge subcell.

| **Subcell** | **Voc (V)** | **Vmp (V)** | **Radiative fraction $\eta_{RAD}(V_{mp})$ (%)** |
|---|---|---|---|
| GaInP | 1.35 | 1.12 | 0.5 |
| GaInAs | 0.95 | 0.84 | 0.5 |
| Ge | 0.29 | 0.22 | 2 |

**Table 6**. Subcell radiative fraction of the total recombination current at their respective maximum power points for the triple junction device showing overall strong non-radiative domaince.

X.4.5 Conclusions

The fundamental loss mechanisms we have seen can be addressed by bandgap engineering, and via the flexible range of materials available in the III-V family. However, materials limitations still impose compromise. These include flexible design at the expense of material purity (the metamorphic route), or greater material quality at the expense of design (homomorphic lattice matched design).

A study of single, tandem, and triple junctions shows impressively increased efficiencies, which are tantalisingly close to the radiative limit. However, increasing the number of junctions brings a diminishing rate of efficiency improvement along the lines we've seen in the earlier discussion on thermalisation losses. In addition, practical issues move the design further from the radiative limit, due partly to increasingly complex manufacturing and lower materials quality.

These considerations are best summarised by the results under one sun AM1.5G illumination showing a manufactured tandem operating at 71% of the radiative limit, versus 66% for a triple junction cell: figures that show potential for improvement.

On that note, we find a comparable triple junction cell has achieved 40.7% under concentrated sunlight.[44] Further triple junction results from Stan *et al.*[45] are a sign that, as mentioned above, there is scope for significant improvement with these high efficiency strategies, among which is the trend towards radiative dominance.

A final point to note is that the simple use of increased light concentration allows operation in a regime closer to radiative dominance – as long as the cell material is sufficiently pure to deliver a ideality n=1 that is explicitly dominated by radiative recombination, as opposed to one dominated by non-radiative Shockley injection recombination pathways in the charge neutral layers.

X.5 Remarks on nanostructures

The bulk semiconductor structures described so far show significant scope for improvement as we have shown on a technological front. Other approaches using the flexibility of the III-V materials have more fundamental potential going beyond the limitations of *pn* multi-junctions devices.

The first key concept as we saw earlier remains the equilibrium population of majority and minority carriers leading to thermalisation loss, and the delivery of all carriers at a single potential. A second key concept is the symmetry of bulk materials resulting in spatially homogeneous properties, such as homogeneous emission over all solid angles. Lifting this homogeneity as mentioned earlier allows reduction of radiative losses for example, and opens the possibility of further gains by reducing structural symmetry. This question of symmetry overlaps with the concept of meta-materials in the general sense: geometric arrangement of available materials such as to modify their combined properties.

To conclude this chapter we will now mention briefly some concepts in III-V solar cell research touching on these issues, and with a common theme which is the modification of bulk materials properties by manipulating materials and geometries on the nano scale, and thereby creating spatially inhomogeneous materials properties.

A design addressing these issues is the hot carrier cell[46] which uses two concepts of slowed carrier thermalisation and energy-selective carrier extraction. The thermalisation rate is decreased by phonon emission rate reduction as a result of modifying the phonon density of states, for example with the use of nanostructures on the quantum scale in two[47] or three dimensions: Quantum dots (QDs) and quantum wells (QWs).[48] The second is achieved by modifying the carrier density of states at or near the contacts and allowing only a narrow energy spectrum of carriers being transported to the contacts. Again, this is achievable by structures that provide well defined energy bands, which are, again, QWs and QDs .

This mention of hot carrier cells, and manipulating carrier energy distributions sharply emphasises the promise of quantum confined structures and the relevance of III-V materials to this field.  There is a rich and fascinating body of research on related issues ranging over quantum wires, quantum dots, and quantum wells.

We will conclude with some comments on quantum well solar cells (QWSCs), a concept that has been developed nearly exclusively in III-V materials for a number of years, and reviewed recently by Barnham.[49] The QWSC is a *pin* structure with lower gap quantum wells sandwiched between higher gap barriers in the undoped intrinsic *i* region  and higher gap doped p and n layers It was initially proposed and studied in the AlGaAs materials system  as a means of extending the absorption of a solar cell whilst keeping a junction potential and hence a $V_{OC}$ determined largely by the higher gap bulk regions enclosing the quantum wells. It transpires, however, that the loss mechanisms mentioned above, and ultimately the fundamental Planck radiative efficiency limit, remain determined largely by the well material, that is, the lower bandgap.

The design has led to a lively debate summarised in the recent reference[49] and a number of phenomena going beyond the bulk semiconductor operating regime. The first is signs of reduced QFL separation in the wells implying decreased carrier populations and decreased recombination relative to bulk samples. Secondly, luminescence studies have shown signs of hot carrier populations in the wells[51]. Finally, however, these effects have not, to date, lead to verifiable operation in efficiency regimes going beyond the bulk.

On the materials front, practical materials advantages of the QWSC have been identified. The first and most practical is the development of the strain balancing technique, whereby alternating quantum well and barrier layers, typically made of GaInP and GaInAs, are grown lattice mismatched on a GaAs substrate but with thicknesses below the critical thickness. The alternating tensile and compressive strain in wells and barriers leads to strain balancing with no net generation of defects. An arbitrary number of quantum well / barrier periods may be grown in principle. The resulting design is the SBQWSC or strain balanced quantum well solar cell.

A second and more fundamental advance is the operation of these cells in the radiative recombination limit.[51,52] The recombination in this design may be up to 90% radiatively dominated as a consequence of the lower quantum well gap being located in the high purity, high mobility $i$ region as we discussed earlier in this chapter.

As a consequence of this radiative dominance, the design symmetry may be further manipulated to restrict emission by fabricating mirrors on all surfaces oriented away from the incoming solar spectrum (essentially the back surface of the cell). The fundamental difference between this approach applied to a bulk cell and to the SBQWSC is the absorption range of the bulk charge neutral layers encasing the space charge region: for the bulk cell, luminescence emitted, and reflected from the back cell is partly re-absorbed by the charge neutral layers, and some of this is lost via non-radiative recombination according to fractions as calculated earlier in this chapter; for a SBQWSC on the other hand, the doped, and therefore lossy charge-neutral layers are transparent to the dominant radiative recombination loss because of their greater bandgap. As a result, the non-radiative loss pathway for these photons via the charge-neutral layers is cut off. The only remaining pathways for the luminescence are reabsorption in the radiatively dominated quantum wells or re-emission towards the source of incident radiation, that is, the sun.

This scenario represents the closest achievable design to the fundamental efficiency limit whereby the only recombination loss is radiative emission, retricted to the solid angle of acceptance of the incoming radiation.

On a practical level, a detailed analysis[52] demonstrates intrinsic radiative dominance in QWSCs as opposed to non-radiative dominance in bulk cells. The same work further reveals the interesting switch in behaviour in QWSCs coated with back-surface mirrors. The dark current (and $V_{OC}$) of the QWSC is dominated by the radiative low-gap quantum well layers. The dark current of the mirror-backed cell, however, is dominated by non-radiative recombination in the higher-gap charge-neutral bulk regions of the cell.

To place this in context, results by Quantasol (http://www.jdsu.com/go/quantasol) have achieved a world record efficiency of 28.3% under concentration[53] and other unpublished results have since achieved efficiencies over 40% for multijunction QWSCs since Quantasol moved into management by JDSU.

To conclude this overview of future concepts, these record nanostructured efficiencies illustrate some of the more fundamental routes forwards in solar cell research, which bring together known concepts of multijunction solar cells together with novel physical concepts of heat and light management. In both these concepts, III-V materials remain key for the flexibility of design this family of materials allows.

X.6 Conclusions

Ongoing changes in global energy supply have moved III-V materials from niche applications towards the mainstream. The principal reasons for this are fossil fuel supply, volume fabrication reduction in costs, and increasing cell efficiency, due to the greater flexibility of cell design possible. These benefits, together with concentrating photovoltaics, justify the complex fabrication processes.

In order to appreciate the relevance of flexible materials, an investigation of high efficiency strategies is necessary. We find that fundamental light and thermal management are key, which analytical analysis allows us to explicitly quantify, together with other materials-related loss mechanisms. An analysis of bulk III-V *pn* and *pin* solar cells shows interesting and contrasting contributions from radiative and non-radiative losses, and how closely the cells approach the radiative limit. The analysis emphasises that bulk single-gap cells are inherently non-radiatively dominated. An analysis of record multijunction cells gives understanding into how thermal management has increased efficiency to date. The relatively low radiative efficiency in these designs, however, emphasises that there is significant potential for improvement.

A brief note on high efficiency nanostructured concepts, some on the verge of commercial success, underline the relevance of the III-V materials family for future concepts. As time goes by and energy costs inevitably increase, the importance of these materials seems set to remain central to development of sustainable energy supply.

**References**

[1] VM Andreev, "GaAs and High-Efficiency Space Cells" in "Practical handbook of photovoltaics: fundamentals and applications", Eds. T. Markvart and Luis Castañer, Elsevier, (2003).

[2] Gregory Phipps, Claire Mikolajczak and Terry Guckes, Renewable Energy Focus

Volume 9, Issue 4, July-August 2008, Pages 56, 58-59.

[3] S. M. Sze and Kwok K. Ng, Physics of Semiconductor Devices, (7th edition, John Wiley & Sons, 2007).

[4] T.P.Pearsall, GaInAsP Alloy Semiconductors, John Wiley and Sons (1982).

[5] S.Adachi, J. Appl. Phys., 66, no.12, 6030-6040 (1989).

[6] I. Vurgaftman, J. R. Meyer, and L. R. Ram-Mohan, J. Appl. Phys. 89, 5815 (2001)

[7] Ottfried Madelung Semiconductors: Data Handbook, 3rd edition Springer (2004).

[8] D.E.Aspnes, Phys. Rev., B14, no.12, pp.5331-5343 (1976)

[9] P Bhattacharya, Semicond. Sci.Technol. 3 pp. 1145-1156 (1988).

[10] Sadao Adachi (Ed.), Properties of Aluminium Gallium Arsenide, EMIS datareviews series no. 7, INSPEC (1993).

[11] J. C. Bourgoin, S. L. Feng, H. J. von Bardeleben, Phys. Rev. B Vol. 40, 11, pp. 7663-7670 (1989).

[12] V. M. Andreev, V. P. Khvostikov, V. R. Larionov, V. D. Rumyantsev, E. V. Paleeva, M. Z. Shvarts, SEMICONDUCTORS, Vol. 33, 9, p. 976 (1999).

[13] Wei Zhang, Mengyan Zhang, Mingbo Chen, Depeng Jiang, Liangxing Wang, Proc. ISES WORLD CONGRESS 2007 (VOL. I – VOL. V), 4, 996-999 (2009).

[14] D.J. Friedman*, J.F. Geisz, S.R. Kurtz, J.M. Olson, Journal of Crystal Growth 195 (1998) 409—415

[15] J F Geisz and D J Friedman, Semicond. Sci. Technol. 17 (2002) 769–777

[16] Vijit Sabnis, Homan Yuen, and Mike Wiemer, 7th international conference on concentrating photovoltaic systems CPV7 in press (2012)

[17] Martin A. Green, Keith Emery, Yoshihiro Hishikawa and Wilhelm Warta, Prog. Photovolt: Res. Appl.; 17:320–326 (2009).

[18] H.-J. Schimper, Z. Kollonitsch, K. Möller, U. Seidel, U. Bloeck, K. Schwarzburg, F. Willig, T. Hannappel, proc. 20th EUPVSEC, 6 – 10 June 2005, Barcelona, Spain, pp. 492-494

[19] F.Himrane, N.M. Pearsall, R.Hill, IN: IEEE Photovoltaic Specialists Conference, 18th, Las Vegas, NV, October 21-25, 1985, Conference Record (A87-19826 07-44). New York, Institute of Electrical and Electronics Engineers, Inc., p. 338-343 (1985).

[20] Pearsall, T.P. Ed. Properties, Processing and Applications of Indium Phosphide, IEE/INSPEC,The Institution of Electrical Engineers, London, (2000).

[21] R. A. Stradling and P. C. Klipstein, Growth and Characterisation of semiconductors, (Adam Hilger, Bristol and New York, 1990).

[22] C. Kittel, Introduction to Solid State Physics, 7th Ed., Wiley, New York, (1996).

[23] R. R. King, D. C. Law, K. M. Edmondson, C. M. Fetzer, G. S. Kinsey, H. Yoon, R. A. Sherif, and N. H. Karam, Appl. Phys. Lett. 90, 183516 (2007).

[24] S. M. Hubbard, C. D. Cress, C. G. Bailey, R. P. Raffaelle, S. G. Bailey, and D. M. Wilt, Appl. Phys. Lett. 92, 123512 (2008).

[25] M. Hermle, G. Létay, S. P. Philipps and A. W. Bett, Prog. Photovolt: Res. Appl.; 16:409–418 (2008).

[26] Alex Trellakis · Tobias Zibold · Till Andlauer · Stefan Birner · R. Kent Smith · Richard Morschl · Peter Vogl, J Comput Electron 5:285–289 (2006).

[27] W. Shockley "Electrons and Holes in Semiconductors", D. Van Nostrand, Princeton, N.J., 1950.

[28] Matthias E. Nell and Allen M. Barnett, IEEE Trans. Elect. Devices, Vol ED-34, No. 2, 1987.

[29] Jenny Nelson The Physics of Solar Cells, Imperial College Press, 2003.

[30] W.Shockley and W.T.Read, Phys. Rev. 87 (1952) 835, and R.N.Hall, Phys. Rev. 87, 387. (1952).

[31] G. L. Araujo and A. Marti, Sol. Energy Mater. Sol. Cells 33, 213, 1994.

[32] Jenny Nelson, Jenny Barnes, Nicholas Ekins-Daukes, Benjamin Kluftinger, Ernest Tsui, and Keith Barnham, J. Appl. Phys., 82, 6240 (1997).

[33] J.P. Connolly, I.M. Ballard, K.W.J. Barnham, D.B. Bushnell, T.N.D. Tibbits, J.S. Roberts, Proc. 19th EUPVSEC, Paris, France, June 2004, pp. 355-359. Preprint: arXiv:1006.1835v2 [cond-mat.mes-hall]

[34] C.H. Henry, J. Appl. Phys 51(8), 4494. (1980).

[35] S. R. Kurtz, J. M. Olson, A. Kibbler, Proc. 21st lEEE PVSC, pp. 138-140 (1990).

[36] T. Trupke, M.A. Green and P. Wuerfel, Improving solar cells by down conversion of high energy photons, J. Appl. Phys. 92 (3) (2002) 1668–1674.

[37] T. Trupke, M.A. Green and P. Wuerfel, Improving solar cells by the upconversion of sub-band-gap light, J. Appl. Phys., 92 (7) (2002) 4117–4122.

[38] A.G. Imenes, D.R. Mills, Solar Energy Materials & Solar Cells 84 (2004) 19–69

[39] Leo Esaki, IEEE Transactions on Electron Devices, vol. ed-23, no. 7, july 1976

[40] T. A. Demassa and D. P. Knott, "The Prediction of Tunnel Diode Voltage-Current Characteristics," Solid-State Electron., 13, (1970), p.131.

[41] D. J. Friedman, Sarah R. Kurtz, K. A. Bertness, A. E. Kibbler, C. Kramer, and J. M. Olson, proc. 1st WCPEC, Hawaii (1994), p. 1829.

[42] Tatsuya Takamoto, Eiji Ikeda, and Hiroshi Kurita, Masamichi Ohmori, Appl. Phys. Lett. 70 (3) (1997)

[43] K.A. Emery, D. Myers and S. Kurtz, Proc. 29th lEEE PVSC, 2002, pp. 840-843.

[44] R. R. King, D. C. Law, K. M. Edmondson, C. M. Fetzer, R. A. Sherif, G. S. Kinsey, D. D. Krut, H. L. Cotal, and N. H. Karam, Proc. 4th WCPEC, (2006) Hawaii, p. 760

[45] M. Stan, D. Aiken, B. Cho, A. Cornfeld, J. Diaz, V. Ley, A. Korostyshevsky, P. Patel, P. Sharps, T. Varghese, Journal of Crystal Growth 310 (2008) 5204–5208

[46] Ross RT, Nozik AJ., J.Appl. Phys, 53 (1982) 3813–3818.

[47] B. K. Ridley J. Phys C: Solid State Phys., 15 (1982), 5899-5917

[48] Santosh K. Shrestha , Pasquale Aliberti, Gavin J. Conibeer, Solar Energy Materials & Solar Cells 94 (2010) 1546–1550

[49] K.W.J. Barnham, I.M. Ballard, B.C. Browne, D.B., Bushnell, J.P. Connolly, N.J. Ekins-Daukes, M. Führer, R. Ginige, G. Hill, A. Ioannides, D.C. Johnson, M.C. Lynch, M. Mazzer, J.S. Roberts, C. Rohr and T.N.D. Tibbits, Chapter 5, in: Nanotechnology for Photovoltaics, ed: Loucas Tsakalakos, CRC Press, (2010).

[50] J. P. Connolly, D.C. Johnson, I.M. Ballard, K.W.J. Barnham, M. Mazzer, T.N.D Tibbits, J.S. Roberts, G. Hill, C. Calder (2007), Proc. ICSC-4, (2007) pp.21-24. arXiv:1006.0660v2 [cond-mat.mes-hall]

[51] K.W.J. Barnham, B.C. Browne, J.P. Connolly, J.G.J. Adams, R.J. Airey, N.J. Ekins-Daukes, M. Führer, V. Grant, K.H. Lee, M. Lumb, M. Mazzer, J.S. Roberts, T.N.D. Tibbits, Proc. Joint 25[th] EUPVSC-5th WCPEC, Valencia (2010), p.234.

[52] James P. Connolly, Chapter 5, in: Advanced Solar Cell Materials, Technology, Modeling, and Simulation, eds: Laurentiu Fara and Masafumi Yamaguchi, IGI Global (2012).

[53] Record verified at Frauenhofer ISE testing centre and reported in the press for example http://www.nanotech-now.com/news.cgi?story_id=33957